\documentclass[pre,aps,showpacs,amsmath,amssymb,superscriptaddress,twocolumn,longbibliography]{revtex4-1}\usepackage{color}
\usepackage{bm}
\usepackage{graphicx}
\usepackage{braket}
\usepackage[plainpages=false,pdfpagelabels,colorlinks=true,linkcolor=red,urlcolor=blue,citecolor=blue,pdftitle={Title},pdfauthor={},pdfdisplaydoctitle=true,pdfduplex=DuplexFlipLongEdge]{hyperref}

\newcommand\ba{\begin{eqnarray}}
\newcommand\ea{\end{eqnarray}}
\newcommand\be{\begin{equation}}
\newcommand\ee{\end{equation}}

\begin{document}
\title{Hellmann Feynman Theorem in Non-Hermitian system}
\author{Gaurav Hajong}
\email{gauravhajong730@gmail.com}
\affiliation{ Department of Physics, Banaras Hindu University, Varanasi-221005, India}
\author{Ranjan Modak}
\email{ranjan@iittp.ac.in}
\affiliation{ Department of Physics, Indian Institute of Technology Tirupati, Tirupati, India~517619}
\author{Bhabani Prasad Mandal}
\email{bhabani@bhu.ac.in}
\affiliation{ Department of Physics, Banaras Hindu University, Varanasi-221005, India}
\begin{abstract}

We revisit the celebrated Hellmann-Feynman theorem (HFT) in the PT invariant non-Hermitian quantum physics framework. We derive a modified version of HFT by changing the definition of inner product and explicitly show that it holds good for both PT broken, unbroken phases and even at the exceptional point of the theory. The derivation is extremely general and works for even PT non-invariant Hamiltonian.  
We consider several examples of discrete and continuum  systems to test our results. We find that if the eigenvalue goes through a real to complex transition as a function of the Hermiticity breaking parameter, both sides of the modified HFT expression diverge at that point. If that point turns out to be an EP
of the PT invariant quantum theory, then one also sees the divergence at EP. Finally, we also derive a generalized Virial theorem for non-Hermitian systems using the modified HFT, which potentially can be tested in experiments.  
\end{abstract}

\maketitle

\section{Introduction}
The so-called Hellmann Feynman Theorem  (HFT) states that the derivative of the system's total energy with respect to a parameter is equal to the expectation value of the derivative of the Hamiltonian of the system with respect to the same  parameter\cite{PhysRev.56.340, 10.1119/1.15842}, which is mathematically written as
\be \label{hft}\frac{\partial E_\lambda}{\partial\lambda}=\Big\langle\psi_\lambda \Big | \frac{\partial \widehat{H}_\lambda}{\partial\lambda}\Big | \psi_\lambda\Big\rangle,
\ee
where $\lambda $ is some arbitrary parameter in the Hamiltonian. The origin of this theorem is not very clear.  Neither Feynman nor Hellmann was the first to use it. Paul Guttinger may have been the first to derive the Eq. (\ref{hft}) in 1932 \cite{1932ZPhy...73..169G}. Hellmann first proved the theorem in 1937 \cite{hellmann1937}.
Later in 1939, Feynman, who apparently  did not know about the earlier works, derived the theorem in his undergraduate thesis and used it to calculate 
forces in molecules directly\cite{PhysRev.56.340}.
 However, the usefulness of HFT in evaluating the expectation  values of dynamical 
quantities in some potential problems have been well demonstrated by several groups\cite{fitts_1999}.
Since its inception, this theorem is widely in use in various branches of physics  and chemistry,  including high energy physics \cite{FERNANDEZ2022169158,PhysRevD.101.094508,PhysRevD.107.094010,PhysRevD.101.054007,sen2021hadronic,PhysRevLett.116.172001,PhysRevD.102.114505,PhysRevD.107.054503,PhysRevD.83.114506,PhysRevD.96.014504,PhysRevD.106.034006,batelaan2023quasidegenerate,hannafordgunn2020scaling,Lacour_2010,can2022compton}, condensed matter physics\cite{PhysRevB.100.024401, 10.1063/5.0053177, PhysRevB.107.195150, De_Rosi_2023, PhysRevB.106.085108, PhysRevB.107.165157, PhysRevB.104.L100506, etea2023study, KARACA2023115674} , machine learning \cite{doi:10.1021/acs.chemrev.0c01111}, quantum chemistry \cite{liu2020pseudomass}.
Various issues and discussions regarding its validity for degenerate systems can be found in \cite{fernandezdegen}.
Our work aims to explore the hidden power of this theorem, particularly to study its applicability beyond the usual quantum framework. 

 During the past two and half decades, non-Hermitian physics has become very exciting and secured its position in frontier research in almost all branches of physics~\cite{rmp_nh,rmp_topology}. Non-Hermiticity can be  originated from exchanges of energy or particles with an environment and leads to rich properties in quantum dynamics~\cite{nh_1,nh_2,nh_3,nh_4}, localization~\cite{mbl_1,mbl_2,pal2022dna},  and topology~\cite{top_1,top_2,top_3,top_4}. On the other hand, non-hermitian Hamiltonian also plays a very important role in understanding quantum measurement problems~\cite{dhar_15,modak2023non}.

 While in general non-Hermitian Hamiltonians have complex eigenvalues, by replacing the self-adjoint condition on the Hamiltonian with a much more  physical and rather less constraining condition of space-time reflection symmetry known as PT-symmetry~\cite{Bender_2002, doi:10.1080/00107500072632, KHARE200053}, can have real eigenvalues. Such systems described by PT-invariant non-Hermitian Hamiltonians can typically be divided into two categories, one in which the  eigenvectors respect PT symmetry and the entire spectrum is real, and the other in which  the whole spectrum or a part of it is complex and the eigenvectors do not respect the PT symmetry.  These are known as PT-unbroken and broken phases, respectively and has been studied extensively in literature\cite{pal2022dna,top_1,top_2,PhysRevA.103.062416,RAVAL2019114699,pseudo1,MANDAL2015185, MANDAL20131043}. The  phase transition point is known as the exceptional point (EP). It has been demonstrated that a consistent quantum theory with an entirely real spectrum, a complete set of orthonormal eigenfunctions having positive definite norms and unitary time evolution in the unbroken phase, can be constructed in a modified Hilbert space equipped with an appropriate  positive definite inner product~\cite{nogo,un_23}. This field of PT-symmetric non-Hermitian physics received a huge boost when the consequence of PT transition was observed experimentally in various analogous systems~\cite{exp_pt_1,exp_pt_2,exp_pt_3,exp_pt_4,exp_pt_5,exp_pt_6}.

 In this work, our main goal is to derive a Hellmann-Feynman-type Theorem 
 for non-Hermitian systems. Given HFT for Hermitian systems has huge applications in different branches of physics, a similar theorem for non-Hermitian systems can also be extremely useful, especially in the context of  open quantum systems. HFT can be used to derive a generalized Virial theorem for quantum particles  with zero-range or finite-range interactions in an arbitrary external potential~\cite{virial}. In the case of Unitary gas in a harmonic trap, this theorem provides us with a relation between the energy of the system and the trapping energy. Virial theorem for such systems has been verified experimentally in cold atom experiments~\cite{virial_exp1,virial_exp2}. We construct the generalized Virial theorem for non-Hermitian systems, which we believe can be tested in experiments. 
 
 The manuscript is organized as follows. First, we derive the modified HFT in Sec. II. Next, we show the validation of it for discrete models and continuum models in Sec.III and Sec. IV respectively. In Sec.V, we prove the generalized Virial theorem for the non-Hermitian system, and finally, we conclude in Sec. VI.

\section{HFT for Non-Hermitian system}

First, let us consider a general PT-symmetric non-Hermitian,
$\widehat{\mathcal{H}}^\dag\not=\widehat{\mathcal{H}}\ \  \big[PT,\widehat{\mathcal{H}}\big]=0$ system. Such systems are characterized by right eigenvectors $|R_i\rangle $ and left eigenvectors 
$|L_i\rangle$ as defined by
$$\widehat{\mathcal{H}}|R_i\rangle=E_i|R_i\rangle \ \ \  \widehat{\mathcal{H}}^\dag |L_i\rangle=E^*_i|L_i\rangle.  $$
Theses eigenvectors form a complete  bi-orthogonal set \cite{shi2009recovering, kleefeld2009construction}, satisfying, $\langle{L_i}|R_j\rangle =\delta_{ij}$ and $\sum_i|R_i\rangle\langle{L_i}| =I $. Alternatively, we can  introduce a Hermitian metric operator G such that $\langle{L_i}|=\langle{R_i|G}$ and use it to define a more general inner product or G-inner product \cite{PhysRevA.100.062118, PhysRevResearch.3.013015}. The orthonormality and completeness relations then are expressed in terms of the G-inner product as
 \be \label{ortho_com_rel}\langle{R_i}|R_j\rangle_G =\langle{R_i}|G|R_j\rangle = \langle{L_i}|R_j\rangle= \delta_{ij} \ee
 The G-operator can be calculated for the theory as, \be \label{G_operator} G= \sum_i |L_i\rangle\langle L_i| = \left [ \sum_i |R_i\rangle\langle R_i|\right ]^{-1}\ee

The expectation value of an observable $O$ will now be defined with respect to the G-inner product as, 
\be \label{G_expt}\langle O\rangle_G=\langle{R_i}|GO|R_i\rangle= \langle{L_i}|O|R_i\rangle\ee

It has been demonstrated  explicitly in Ref.~\cite{un_23,nogo}  that 
$\langle O\rangle_G$ is a real number for any state in the Hilbert space if and only if $O$ satisfies the following condition i.e. 
$$O^\dag G = G O.$$
The observables which obey the above condition are called ``good observables". 

    Now we are in a position to obtain HFT for PT invariant  non-Hermitian quantum mechanics. Let us first concentrate on  the unbroken phase of the theory where Hamiltonian is good observable~\cite{un_23}, i.e.  
\begin{equation}
    \label{good_observable_condition}\widehat{\mathcal{H}}^\dag{G}=G\widehat{\mathcal{H}}
\end{equation}
and hence energy eigenvalues are real, $E_i =E_i^*$. Further, we consider our Hamiltonian depends on a real parameter $\lambda$ and $E_\lambda$ is the energy eigenvalue of an arbitrary right eigenstate 
$|R\rangle$ (note that we drop the suffix $i$ from Eq.~\eqref{ortho_com_rel} to simplify the notation). Differentiating the expression 
$\langle{L}|\widehat{\mathcal{H}}|R\rangle=E_\lambda$ with respect to $\lambda$, we obtain
%$$\pdv{\langle{L}|\widehat{\mathcal{H}}|R\rangle}=\frac{\partial E_\lambda}{\partial\lambda}$$
 \ba
 \frac{\partial E_\lambda}{\partial\lambda}  =&& \Bigl(\frac{\partial{\langle{L}|}}{\partial\lambda}\Bigr)\widehat{\mathcal{H}}\Big|R\Big\rangle+\Big\langle{L}\Big|\frac{\partial{\widehat{\mathcal{H}}}}{\partial\lambda}\Big|R\Big\rangle+\Big\langle{L}\Big|\widehat{\mathcal{H}}\Big(\frac{\partial{|R\rangle}}{\partial\lambda}\Big)\nonumber\\
  =&&E_\lambda\Bigl(\frac{\partial\langle{L}|}{\partial\lambda}\Bigr)\Big|R\Big\rangle+\Big\langle{L}\Big|\frac{\partial\widehat{\mathcal{H}}}{\partial\lambda}\Big|R\Big\rangle+\Big\langle{R}\Big|G\widehat{\mathcal{H}}\Big(\frac{\partial{|R\rangle}}{\partial\lambda}\Big)\nonumber \\
  =&& E_\lambda\Bigl(\frac{\partial\langle{L}|}{\partial\lambda}\Bigr)\Big|R\Big\rangle+\Big\langle{L}\Big|\frac{\partial\widehat{\mathcal{H}}}{\partial\lambda}\Big|R\Big\rangle \nonumber  \\ && +\Big\langle{G^{-1}\widehat{\mathcal{H}}^\dag{G}R}\Big|G\Big(\frac{\partial{|R\rangle}}{\partial\lambda}\Big)\nonumber\\
   \ea
The last term  in the above equation can be simplified by using good observable conditions for the Hamiltonian in the unbroken phase  to obtain
\ba
 \frac{\partial E_\lambda}{\partial\lambda}  =&& E_\lambda\Bigl(\frac{\partial\langle{L}|}{\partial\lambda}\Bigr)\Big|R\Big\rangle+\Big\langle{L}\Big|\frac{\partial\widehat{\mathcal{H}}}{\partial\lambda}\Big|R\Big\rangle+E_\lambda\Big\langle{L}\Big|\Big(\frac{\partial{|R\rangle}}{\partial\lambda}\Big)\nonumber\\ 
 =&&E_\lambda\frac{\partial\langle{L}|R\rangle}{\partial\lambda}+\Big\langle{L}\Big|\frac{\partial\widehat{\mathcal{H}}}{\partial\lambda}\Big|R\Big\rangle \nonumber \\
 \ea
This leads to the HFT in PT-symmetric non-Hermitian (unbroken phase) as 
\begin{equation}
\label{hft_unbroken_1}
\boxed{\Bigl\langle{L}\Big|\frac{\partial\widehat{\mathcal{H}}}{\partial\lambda}\Big|R\Bigr\rangle=\frac{\partial E_\lambda}{\partial\lambda}}
\end{equation}
%\end{document}
or alternatively,
\begin{equation}
\label{hft_unbroken_2}
\boxed{\Bigl\langle{R}\Big|G\frac{\partial\widehat{\mathcal{H}}}{\partial\lambda}\Big|R\Bigr\rangle=\frac{\partial E_\lambda}{\partial\lambda}}
\end{equation}
Eqs.~\eqref{hft_unbroken_1} and \eqref{hft_unbroken_2} represent the modified HFT in the unbroken region.

Now we show that the same expression also holds in the  broken phase. We take a different approach here as the Hamiltonian is not a good observable in the broken phase.

Let us start differentiating the complex $E_\lambda = \langle L|\widehat{\mathcal{H}}|R\rangle$ with respect to the parameter $\lambda$
\ba
\frac{\partial E_\lambda}{\partial\lambda} =&& \Bigl(\frac{\partial\langle{L}|}{\partial\lambda}\Bigr)\widehat{\mathcal{H}}\Big|R\Big\rangle+\Big\langle{L}\Big|\frac{\partial\widehat{\mathcal{H}}}{\partial\lambda}\Big|R\Big\rangle+\Big\langle{L}\Big|\widehat{\mathcal{H}}\Big(\frac{\partial|R\rangle}{\partial\lambda}\Big)\nonumber\\
=&& E_\lambda\Bigl(\frac{\partial\langle{L}|}{\partial\lambda}\Bigr)\Big|R\Big\rangle+\Big\langle{L}\Big|\frac{\partial\widehat{\mathcal{H}}}{\partial\lambda}\Big|R\Big\rangle+\Big\langle{\widehat{\mathcal{H}}^\dag{G}{R}}\Big|\Big(\frac{\partial|R\rangle}{\partial\lambda}\Big). \nonumber \\
~\nonumber\\
\ea

On the other hand,   $\mathcal{H}^\dag|L\rangle=E^*_\lambda|L\rangle$ can be written as 
$\langle\mathcal{H}^\dag{G}R|=E_\lambda\langle{R}|G$. Using this in the last term of the above equation we obtain
\be 
 \frac{\partial E_\lambda}{\partial\lambda}=  E_\lambda\frac{\partial\langle{L}|R\rangle}{\partial\lambda}+\Big\langle{L}\Big|\frac{\partial\widehat{\mathcal{H}}}{\partial\lambda}\Big|R\Big\rangle.
 \ee
 \be\label{hft_broken_eqn} 
 \boxed{\frac{\partial E_\lambda}{\partial\lambda}=  \Big\langle{L}\Big|\frac{\partial\widehat{\mathcal{H}}}{\partial\lambda}\Big|R\Big\rangle=\Big\langle{R}\Big|G\frac{\partial\widehat{\mathcal{H}}}{\partial\lambda}\Big|R\Big\rangle}
 \ee
 This leads to the same form of HFT as in the case of the unbroken phase. Remarkably, this proof is extremely general and should work
 for any non-Hermitian Hamiltonian even if the Hamiltonian is not PT invariant, i.e., $[\widehat{\mathcal{H}},PT]\ne 0$. However, note that given the Hamiltonian itself a good observable in the PT-symmetric phase, the LHS and RHS of the expression of modified HFT Eq.~\eqref{hft_unbroken_1} will be always real, but in the broken phase or for non-PT-invariant systems, they can be complex numbers. 
 
 Next, we will consider explicit examples of non-Hermitian Hamiltonian, discrete and continuum to verify our claim.

\section{Discrete models}
%\end{document}
%\begin{enumerate}
%  \subsection{ Two level system}  

First, we consider a PT-symmetric non-Hermitian two-level system \cite{PhysRevA.103.062416} described by the following Hamiltonian, 
$$H_{2\times2}=
\begin{pmatrix}{i\lambda}&{-1}\\ {-1}&{-i\lambda}.
\end{pmatrix}$$
This system undergoes a PT phase transition at $\lambda_c =1$
The eigenvalues are $E_\pm=\pm\sqrt{1-\lambda^2}$ and corresponding right eigenvectors in the unbroken phase  i.e. $\lambda<1$ are given by,
$$|R_{+}\rangle=\frac{i}{\sqrt{2\cos\alpha}}\begin{pmatrix}{e^{-\frac{i\alpha}{2}}}\\ {- e^{\frac{i\alpha}{2}}}
\end{pmatrix}$$ \\ $$|R_{-}\rangle=\frac{1}{\sqrt{2\cos\alpha}}\begin{pmatrix}{e^{\frac{i\alpha}{2}}}\\ {e^{-\frac{i\alpha}{2}}}
\end{pmatrix}$$ where, $\sin\alpha=-\lambda$. The G-operator in the unbroken phase reads as,
\be G^u=\frac{1}{\sqrt{1-\lambda^2}}\begin{bmatrix}{1}&{i\lambda}\\ {-i\lambda}&{1}
\end{bmatrix}.\ee

Now it is straightforward to check that for $|R_+\rangle$ in the unbroken phase 
\be 
\label{+revur} 
\Big\langle{R_+}\Big|G^u\frac{\partial{H_{2\times2}}}{\partial\lambda}\Big|R_+\Bigr\rangle=-\frac{\lambda}{\sqrt{1-\lambda^2}} = \frac{\partial{E_+}}{\partial\lambda}.
\ee
Even in the broken phase, one obtains, 
\be 
\label{+revur1} 
\Big\langle{R_+}\Big|G^b\frac{\partial{H_{2\times2}}}{\partial\lambda}\Big|R_+\Bigr\rangle=i\frac{\lambda}{\sqrt{1-\lambda^2}} = \frac{\partial{E_+}}{\partial\lambda}.
\ee
where $G^b$ is the $G$ matrix in the broken phase.$|R_-\rangle$ also satisfies Eq.~\eqref{hft_broken_eqn} for both broken and unbroken phases.
This confirms the validity of the modified HFT for this model. 
We have explicitly verified the validity of modified HFT even for  $4 \times 4 $  PT-symmetric non-Hermitian system (see Appendix A for details). 

\begin{figure}[h]   
    \centering\includegraphics[width=0.48\textwidth]{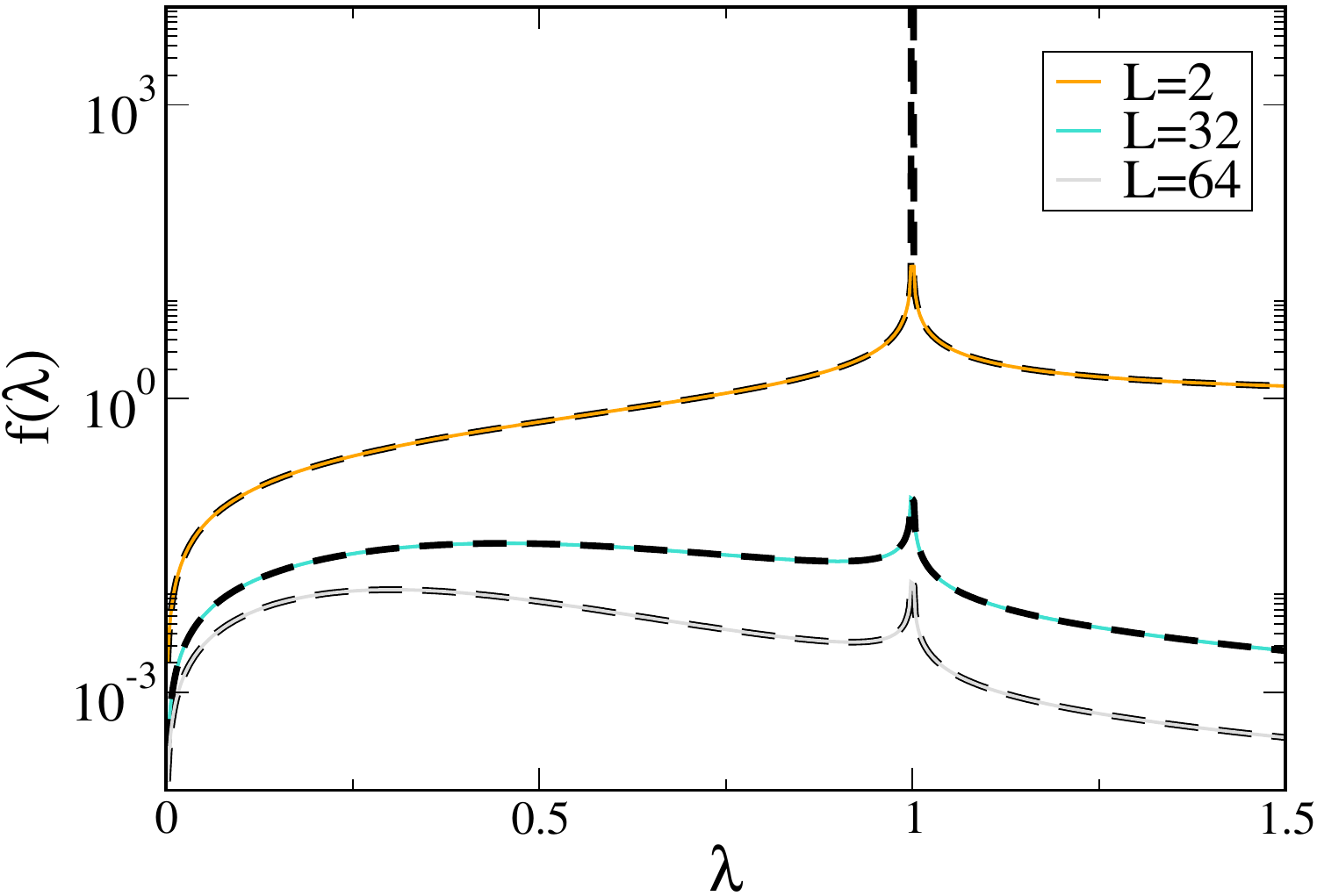}
    \caption{Comparison between the absolute values of LHS  and RHS of the Eq.~\eqref{hft_unbroken_1} for Hamiltonian Eq.~\eqref{hamiltonian_pt} in solid lines and dashed lines respectively for different values of $L$. }
    \label{hft2}
\end{figure}
\begin{figure}[h]   
    \centering\includegraphics[width=0.48\textwidth]{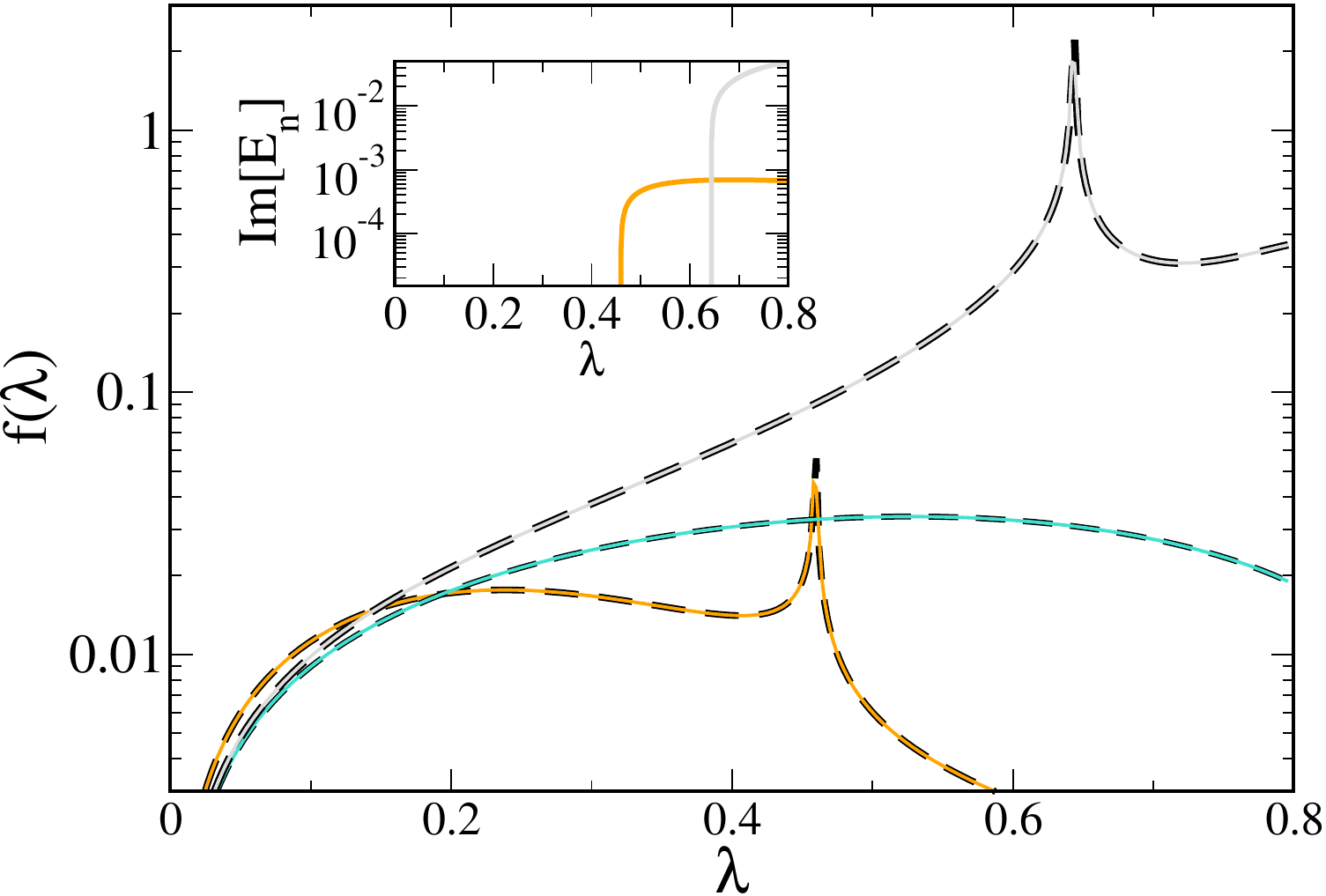}
    \caption{Comparison between the absolute value of LHS  and RHS of the Eq.~\eqref{hft_unbroken_1} for Hamiltonian Eq.~\eqref{hamiltonian_pt4} in solid lines and dashed lines respectively for  $L=64$ and different eigenstates. Inset shows the imaginary part of those eigenvalues as a function of $\lambda$. Note that the cyan color plot is missing from the inset, that is because imaginary part of that eigenvalue is zero for $\lambda\le 0.8$.  }
    \label{hft1}
\end{figure}

Next, we study $L\times L$ version of the above $2\times 2$, such Hamiltonian can be interpreted as a model of  non-interacting fermions in 1D lattice with open boundary and  
described by the following   Hamiltonian, 
\begin{eqnarray}
 {H_0}=-\sum _{j=1}^{L-1}(\hat{c}^{\dag}_j\hat{c}^{}_{j+1}+\text{H.c.}) \nonumber \\
 \label{hamiltonian}
\end{eqnarray}
where $\hat{c}^{\dag}_j$ ($\hat{c}_{j}$) is the fermionic creation (annihilation) operator at site $j$, 
which satisfies standard anti-commutation relations. $L$ is the size of the system, which we set to be an even number for all our calculations ( we choose the lattice spacing as unity). 

To make the Hamiltonian  $PT$ symmetric and non-Hermitian, we add a local  term at site $L/2$ and $L/2+1$. 
The $PT$ symmetric Hamiltonian reads as, 

\begin{eqnarray}
H_{L\times{L}}=H_0+i\lambda (\hat{n}_{L/2}-\hat{n}_{L/2+1}) \nonumber \\
\label{hamiltonian_pt}
\end{eqnarray}

where, $\hat{n}_j=\hat{c}^{\dag}_j\hat{c}_{j}$ is the number operator and $\lambda$ is identified as the Hermiticity breaking 
parameter. While under Parity transformation $c_j \to c_{L-j+1}$, Time reversal symmetry operation changes $i \to -i$. 
Hence, $H_{L\times{L}}$ remains invariant under $PT$ transformation, which implies $[H_{L\times{L}},PT]=0$.  For non-zero values of $\lambda$, $H_{L\times{L}}$ is non-Hermitian. For $L=2$, this model is identical to the two-level system we have studied previously. For any finite and even $L$
the Hamiltonian Eq.~\eqref{hamiltonian_pt} shows a PT transition at 
$\lambda=1$~\cite{PhysRevA.103.062416}. Like a two-level system, all eigenvalues are completely complex for $\lambda>1$. Note that it's not necessary to have all eigenvalues to be complex in the PT broken phase; in contrast, we only need two of them to be complex. Fig.~\ref{hft2} 
shows excellent agreement between LHS and RHS of Eq.~\eqref{hft_unbroken_1} (which we refer to as $f(\lambda)$). Interestingly, while Eq.~\eqref{hft_unbroken_1} remains valid even when we approach
towards EP i.e. $\lambda=1$, but the numerical value seems to diverge at EP. This is already clear for two-level systems from Eqs.~\eqref{+revur} and \eqref{+revur1}, which diverge in the $\lambda\to 1$ limit.  While this result is tempting to conclude that near EP LHS and RHS of Eq.~\eqref{hft_unbroken_1} always diverges, it turns out to be not always true. Next, we study another model that is described by the Hamiltonian,
\begin{eqnarray}
\tilde{H}_{L\times{L}}=H_0+i\lambda \sum_{j=L/2-(r-1)}^{L/2+r}(-1)^j\hat{n}_j ,\nonumber \\
\label{hamiltonian_pt4}
\end{eqnarray}
where we chose $r=2$. Hamiltonian Eq.~\eqref{hamiltonian_pt4}
is interesting in the sense that in the PT broken phase of this model, not all eigenvalues are completely complex. A parameter regime exists when some eigenvalues are completely real, even in the broken phase.  We find that while all eigenvalues and eigenstates modified HFT are valid, the divergence of LHS and RHS of Eq.~\eqref{hft_unbroken_1} near EP occurs only for those eigenstates that show real to complex transition at EP. Eigenstates correspond to eigenvalues that remain real even after EP, for which no divergence has been observed at EP (see Fig.~\ref{hft1}). We focus on three eigenstates of the Hamiltonian Eq.~\eqref{hamiltonian_pt4} for $L=64$ in Fig.~\ref{hft1} and also have plotted imaginary part of those eigenvalues as a function of $\lambda$. Note that while EP of this model  corresponds to $\lambda_c\simeq 0.48$ ~\cite{PhysRevA.103.062416,un_23}, some states show no signature of divergence in $f(\lambda)$ at EP. However, it seems that $f(\lambda)$ tends to diverge when that eigenvalue also shows a real to complex transition. Hence, we conjecture  the divergence of Eq.~\eqref{hft_unbroken_1} is because of the real to complex transition of eigenvalues, and it explicitly has nothing to do with EP. Also, we believe 
that  a variant of our lattice models with gain
and loss is experimentally realizable in an ultracold fermionic system.

%%%%%%%%%%%%%%%%%%%%%%
\section{Continuum Model}
%%%%%%%%%%%%%%%%%%%%%%%%%%%%%

\begin{figure}[h!]   
    \centering\includegraphics[width=0.48\textwidth]{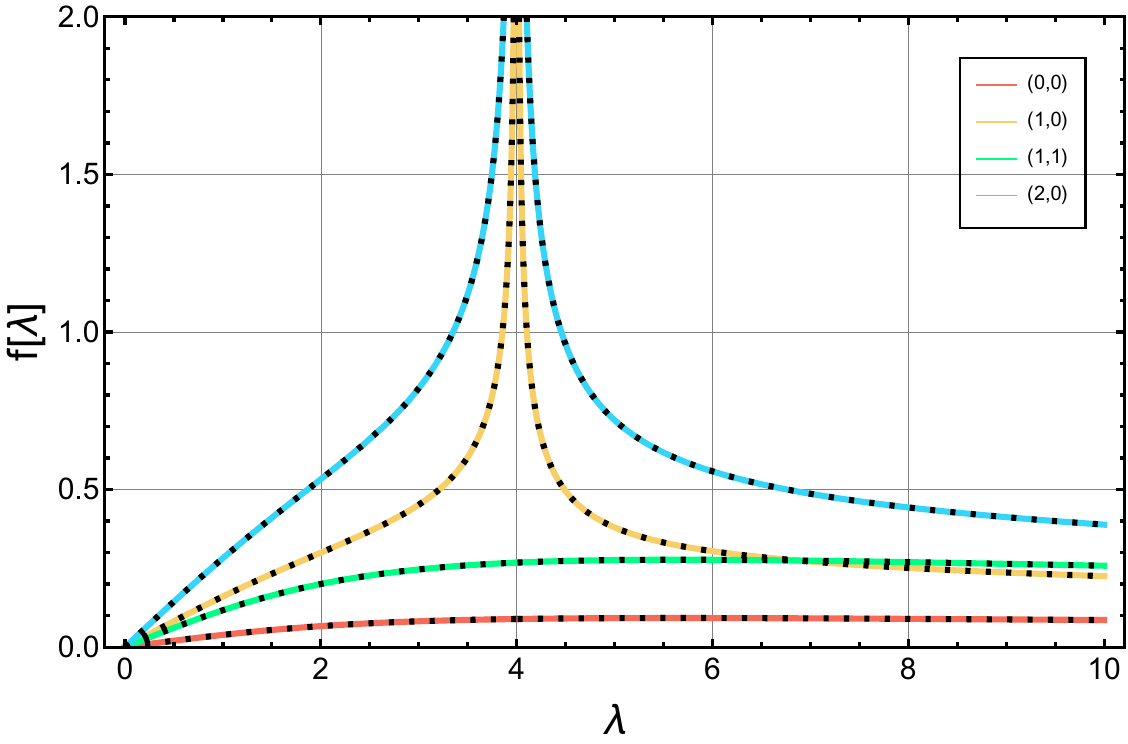}
    \caption{Comparison between the absolute value of $\Big\langle\frac{\partial\widehat{\mathcal{H}}}{\partial\lambda}\Big\rangle_G$ and $\frac{\partial{E_{n_1,n_2}}}{\partial\lambda}$ for both broken and unbroken regions for the states $(n_1,n_2) = (0,0), (1,0), (1,1)$ and $(2,0)$. The  solid lines and black dotted lines indicate the LHS  and RHS of Eq.~\eqref{hft_integral}.  }
    \label{fig:1,0*abo}
\end{figure}

Now we consider the continuum model as an example; we take a 2-d anharmonic oscillator with a  non-hermitian interaction term \cite{MANDAL20131043},  which is described by 

\begin{equation}
\label{pt_hamiltonian}
   H_{2d}=\frac{p_x^2}{2m}+\frac{p_y^2}{2m}+\frac{1}{2}m\omega_x^2x^2+\frac{1}{2}m\omega_y^2y^2+i\lambda{xy},
\end{equation}
where $\lambda$ is real and $\omega_x\neq\omega_y$.
 This system can be solved exactly. The energy eigenvalues and the right eigenvectors are given by, 
  \begin{eqnarray}
\label{right_eigen_vector_n1,n2}
R_{n_1,n_2} &= &Ne^{-\frac{m}{2\hbar}\left[(C_1X^2+C_2Y^2)\right]}H_{n_1}\left(\alpha_1X\right)H_{n_2}\left(\alpha_2Y\right) \nonumber \\
E_{n_1,n_2} &= & (n_1+\frac{1}{2})\hbar{C_1}+(n_2+\frac{1}{2})\hbar{C_2}
\end{eqnarray}
\mbox{where, $\alpha_1=\sqrt{\frac{mC_1}{\hbar}}$ and $\alpha_2=\sqrt{\frac{mC_2}{\hbar}}$}
and the 
left eigenvectors are given by, 
\begin{eqnarray}
\label{left_eigen_vector_n1,n2}
L_{n_1,n_2}=Ne^{-\frac{m}{2\hbar}\left[(C_1^*{X^*}^2+C_2^*{Y^*}^2)\right]}H_{n_1}\left(\alpha_1^*X^*\right)H_{n_2}\left(\alpha_2^*Y^*\right) \nonumber \\
\end{eqnarray}
\mbox{where, $\alpha_1^*=\sqrt{\frac{mC_1^*}{\hbar}}$ and $\alpha_2^*=\sqrt{\frac{mC_2^*}{\hbar}}$}

where, 

$X=\sqrt{\frac{k+1}{2}}x-i\sqrt{\frac{k-1}{2}}y$

$Y=i\sqrt{\frac{k-1}{2}}x+\sqrt{\frac{k+1}{2}}y$.
Moreover, 
$C_1=\sqrt{\frac{1}{2}\Big(\omega_+^2-\frac{\omega_-^2}{k}\Big)},\qquad{C_2=\sqrt{\frac{1}{2}\Big(\omega_+^2+\frac{\omega_-^2}{k}\Big)}}$

$\omega_+^2=\omega_y^2+\omega_x^2,\qquad{\omega_-^2=\omega_y^2-\omega_x^2}$

$\frac{1}{k}=\sqrt{1-\frac{\lambda^2}{\lambda_c^2}}$
$\lambda_c=\frac{m\omega_-^2}{2}.$ 
\\\\
The modified  HFT in the case of continuum models is written in the integral form as,  

\begin{eqnarray}
\label{hft_integral}
\frac{\int{{\Big(L_{n_1,n_2}^*}\Big)\frac{\partial{H_{2d}}}{\partial\lambda}\Big(R_{n_1,n_2}\Big)dxdy}}{\int{{\Big(L_{n_1,n_2}^*}\Big)\Big(R_{n_1,n_2}\Big)dxdy}}=\frac{\partial{E_{n_1,n_2}}}{\partial\lambda}.
\end{eqnarray}
It is straightforward to show that when $n_1= n_2$, the energy eigenvalues are real, and the eigenvectors are PT-symmetric over all values of $\lambda$. We have explicitly shown that the LHS and RHS of Eq.~\eqref{hft_integral} are the same for ground state ( $n_1=n_2=0$) [ see Appendix B for details ].

Next we consider some of the low lying  excited states. The energy eigenvalues can be real or complex depending on the value of $\lambda$. We have plotted the absolute value of LHS and RHS of the modified HFT for all these states in Fig.~\ref{fig:1,0*abo} to check the validity of the theorem. Note that we have also checked explicitly for $(1,0), (2,0) $, the real and imaginary parts of the LHS and RHS of the Eq.~\eqref{hft_integral} separately (see Appendix B). We also find that  $f(\lambda)$ for the  1st excited state  diverges near $\lambda=4$ and  at the same value of $\lambda$ the eigenvalue of the 1st excited state also shows real to complex transition. This result strengthens our previous claim, i.e., the divergence of $f(\lambda)$ corresponds to the real to the complex transition of the eigenvalues. 

\section{Virial theorem for non-Hermitian system}
A generalized virial theorem has been derived for $N$ particles quantum system \cite{virial}, with
arbitrary statistics and dispersion relations. One can consider a general Hamiltonian,
\be
\widehat{\mathcal{H}} = \widehat{\mathcal{H}}' + U(r_1, ...,r_N)
\ee
where $\widehat{\mathcal{H}}'$ and its domain depend on p parameters
$l_1 ,...,l_p$ which have the dimension of a length,  and $U(r_1 ,...,r_N)$ is an arbitrary function, where  $r_i$ is the position of particle $i$. 
Using HFT for the Hermitian system, it has been shown in Ref.~\cite{virial} that for any stationary state energy $E$
the following relation holds, 
\be
E=\left\langle{U+\frac{1}{2}\sum_{i=1}^N{r_i}.\nabla{U\left(r_i\right)}}\right\rangle-\frac{1}{2}\sum_{q=1}^pl_q{\frac{\partial E} {\partial l_q} }.
\ee
If $\widehat{\mathcal{H}}$ is non-Hermitian, it is straightforward to derive a generalized version of the virial theorem using the modified HFT. It just reads as, 
\be
\label{virial_nh}
E=\left\langle{U+\frac{1}{2}\sum_{i=1}^N{r_i}.\nabla{U\left(r_i\right)}}\right\rangle_G-\frac{1}{2}\sum_{q=1}^pl_q{\frac{\partial E} {\partial l_q} }.
\ee
Next, we take a concrete example of 1D non-Hermitian harmonic oscillator with complex angular frequency; the Hamiltonian reads as, 

\be
\label{NH-NPT-symmetric}
H_{1d}=\frac{p^2}{2}+\frac{1}{2}\Omega^2x^2
\ee

where, $\Omega=\omega_1+i\omega_2$ and $\omega_1\ne0$. Note that this Hamiltonian is not PT invariant. However, given our modified HFT holds even for the Hamiltonian, which is not PT invariant, one can still use the modified HFT to derive the generalized virial theorem Eq.~\eqref{virial_nh}. 

The trapping energy for such a non-hermitian system is given by
\be
\label{trapping_energy}
\tilde{E}_{tr}=\frac{1}{2}\left\langle{U\left(x\right)+\frac{1}{2}{x}.\nabla_x{U\left(x\right)}}\right\rangle_G
\ee

%where G signifies the inner product in terms of the G-operator.

We have,
\be
U=\frac{1}{2}\Omega^2x^2
\ee
Therefore, Eq.~\eqref{trapping_energy} implies,
\ba
\label{trap}
\tilde{E}_{tr}=\frac{1}{2}\left\langle{\frac{1}{2}\Omega^2x^2+\frac{1}{2}{x}.\nabla{\left(\frac{1}{2}\Omega^2x^2\right)}}\right\rangle_G\nonumber\\
 =\frac{1}{2}\left\langle{\Omega^2x^2}\right\rangle_G
\ea

Using the definition of the G-inner product given in Eq.~\eqref{G_expt}, we will now deduce Eq.~\eqref{trap} for the ground state and all the excited states. For ground state,\\

Left eigenvector of $H_{1d}$ is given by
\be
L_0={\left(\frac{\omega_1-i\omega_2}{\pi }\right)}^{\frac{1}{4}}e^{-\frac{\omega_1{x^2}}{2}}e^{\frac{i\omega_2{x^2}}{2}}
\ee
and 
right eigenvector of $H_{1d}$ is given by
\be
R_0={\left(\frac{\omega_1+i\omega_2}{\pi }\right)}^{\frac{1}{4}}e^{-\frac{\omega_1{x^2}}{2}}e^{-\frac{i\omega_2{x^2}}{2}}. 
\ee
Therefore, Eq.~\eqref{trap} implies,
\ba
\tilde{E}_{tr_0}=&&\frac{\int{{\Big(L_{0}^*}\Big){\frac{1}{2}\Omega^2x^2}\Big(R_{0}\Big)dx}}{\int{{\Big(L_{0}^*}\Big)\Big(R_{0}\Big)dx}}\nonumber\\
=&&\frac{\int{e^{-\omega_1{x^2}}e^{-i\omega_2{x^2}}{\frac{1}{2}\Omega^2x^2}dx}}{\int{e^{-\omega_1{x^2}}e^{-i\omega_2{x^2}}dx}}\nonumber\\
=&&\frac{1}{4}\Omega=\frac{1}{2}E_0,
\ea
where $E_0$ is the ground state energy of the Hamiltonian Eq.~\eqref{NH-NPT-symmetric}.

One can also compute trapping energy for the $n^{th}$ eigenstate. 
we have,
\be
x=\frac{1}{\sqrt{2\Omega}}\left(a^\dagger+a\right)
\label{x_rel}
\ee
and
\be
p=i\sqrt{\frac{\Omega}{2}}\left(a^\dagger-a\right)
\label{p_rel}
\ee
such that 
\be
a|n\rangle=\sqrt{n}|n-1\rangle \quad \text{and} \quad a^\dagger|n\rangle=\sqrt{n+1}|n+1\rangle
\label{a_adag}
\ee
where, $|n\rangle$ is the $n^{th}$ state of $H_{1d}$, also, the right eigenvector, $|R_n\rangle$ of $H_{1d}$, $\left(i.e., |R_n\rangle=|n\rangle\right)$.

Similarly, 
\be
\label{NH-PT+}
H_{1d}^\dagger=\frac{p^2}{2}+\frac{1}{2}{\Omega^*}^2x^2,
\ee

and we have,
\be
x=\frac{1}{\sqrt{2\Omega^*}}\left(b^\dagger+b\right)
\ee
and
\be
p=i\sqrt{\frac{\Omega^*}{2}}\left(b^\dagger-b\right)
\ee
such that 
\be
b|\bar{n}\rangle=\sqrt{\bar{n}}|\bar{n}-1\rangle \quad \text{and} \quad b^\dagger|\bar{n}\rangle=\sqrt{\bar{n}+1}|\bar{n}+1\rangle
\ee
where, $|\bar{n}\rangle$ is the $\bar{n}^{th}$ state of $H_{1d}^\dagger$, also, the left eigenvector, $|L_n\rangle$ of $H_{1d}$, $\left(i.e., |L_n\rangle=|\bar{n}\rangle\right)$.\\

Therefore, by using  Eq.~\eqref{G_expt}, we can derive the trapping energy for $n^{th}$-eigenstate as,
\ba
\tilde{E}_{tr_n}=&&\langle{L_n}|\frac{1}{2}\Omega^2x^2|R_n\rangle
=\langle{\bar{n}}|\frac{1}{2}\Omega^2x^2|n\rangle\nonumber\\
=&&\frac{1}{2}\Omega^2\left(\frac{1}{2\Omega}\right)\langle{\bar{n}}|{a^\dagger}^2+a^2+a^\dagger a+aa^\dagger|n\rangle\nonumber\\
=&&\frac{\Omega}{4}\langle{\bar{n}}|2n+1|n\rangle \nonumber \\
=&& \frac{1}{2}\left(n+\frac{1}{2}\right)\Omega=\frac{1}{2}E_n,
\ea
where $E_n$ corresponds to eigenenergy of the $n^{th}$ state.
The relation between the energy eigenvalue and the trapping energy is the same even for the  Hermitian harmonic oscillator.  
%%%%%%%%%
%%%%%%%%%
\section{Conclusions}
In this work, we derive the modified HFT for the non-Hermitian system. Our modified HFT works for both PT-invariant and non-invariant systems. The derivation is extremely general. However, given the Hamiltonian is a good observable in the PT-symmetric phase of the PT invariant system, only in that phase the LHS and RHS of the modified HFT in Eq.~\eqref{hft_unbroken_1} guaranteed to be  completely real. Moreover, if the eigenvalue goes through a real to complex transition as a function of the Hermiticity breaking parameter, the LHS and RHS of the modified HFT diverge at that point. If that point is an EP of the PT invariant quantum theory, then one sees the divergence at EP as well; otherwise, both sides of Eq.~\eqref{hft_unbroken_1} can be finite at EP as well. We test our results for different discrete and continuum models; some of those models have already been experimentally realized and have a huge technological application as a quantum sensor~\cite{exp_pt_6}.  

Finally, we also derive a generalized Virial theorem for non-Hermitian systems and show that for a Harmonic oscillator with complex frequency, the system's energy is twice the trapping energy, precisely what one observes for a Hermitian Harmonic oscillator. For the Hermitian system, the trapping potential energy $E_{tr}$ has been computed  from the density profile~\cite{density_exp_1,density_exp_2}, and the released energy $E-E_{tr}$ from a time-of-flight experiment~\cite{tof_1}. Note that recently there has been an experimental proposal to compute the G-inner product for a non-Hermitian system using weak measurement~\cite{weak_19}; we believe the same strategy can be used to compute both energy and trapping energy of the non-Hermitian system; hence experimental verification of Virial theorem for the non-Hermitian system should be possible. Moreover, given that DNA-unzipping transition can be effectively described by a non-Hermitian Hatano-Nelson model ~\cite{pal2022dna,hatano1996localization}, it will be interesting in future to compute the critical force  for DNA-unzipping transition using the modified HFT. 

{\bf Acknowledgements:}
RM acknowledges the DST-Inspire fellowship by the Department of Science and Technology, Government of India, SERB start-up grant (SRG/2021/002152). BPM acknowledges the research grant for faculty under IoE Scheme (Number 6031) of Banaras Hindu University, Varanasi. GH acknowledges the UGC-JRF Fellowship

\bibliography{ref_3}

\begin{thebibliography}{75}%
\makeatletter
\providecommand \@ifxundefined [1]{%
 \@ifx{#1\undefined}
}%
\providecommand \@ifnum [1]{%
 \ifnum #1\expandafter \@firstoftwo
 \else \expandafter \@secondoftwo
 \fi
}%
\providecommand \@ifx [1]{%
 \ifx #1\expandafter \@firstoftwo
 \else \expandafter \@secondoftwo
 \fi
}%
\providecommand \natexlab [1]{#1}%
\providecommand \enquote  [1]{``#1''}%
\providecommand \bibnamefont  [1]{#1}%
\providecommand \bibfnamefont [1]{#1}%
\providecommand \citenamefont [1]{#1}%
\providecommand \href@noop [0]{\@secondoftwo}%
\providecommand \href [0]{\begingroup \@sanitize@url \@href}%
\providecommand \@href[1]{\@@startlink{#1}\@@href}%
\providecommand \@@href[1]{\endgroup#1\@@endlink}%
\providecommand \@sanitize@url [0]{\catcode `\\12\catcode `\$12\catcode
  `\&12\catcode `\#12\catcode `\^12\catcode `\_12\catcode `\%12\relax}%
\providecommand \@@startlink[1]{}%
\providecommand \@@endlink[0]{}%
\providecommand \url  [0]{\begingroup\@sanitize@url \@url }%
\providecommand \@url [1]{\endgroup\@href {#1}{\urlprefix }}%
\providecommand \urlprefix  [0]{URL }%
\providecommand \Eprint [0]{\href }%
\providecommand \doibase [0]{http://dx.doi.org/}%
\providecommand \selectlanguage [0]{\@gobble}%
\providecommand \bibinfo  [0]{\@secondoftwo}%
\providecommand \bibfield  [0]{\@secondoftwo}%
\providecommand \translation [1]{[#1]}%
\providecommand \BibitemOpen [0]{}%
\providecommand \bibitemStop [0]{}%
\providecommand \bibitemNoStop [0]{.\EOS\space}%
\providecommand \EOS [0]{\spacefactor3000\relax}%
\providecommand \BibitemShut  [1]{\csname bibitem#1\endcsname}%
\let\auto@bib@innerbib\@empty
%</preamble>
\bibitem [{\citenamefont {Feynman}(1939)}]{PhysRev.56.340}%
  \BibitemOpen
  \bibfield  {author} {\bibinfo {author} {\bibfnamefont {R.~P.}\ \bibnamefont
  {Feynman}},\ }\bibfield  {title} {\enquote {\bibinfo {title} {Forces in
  molecules},}\ }\href {\doibase 10.1103/PhysRev.56.340} {\bibfield  {journal}
  {\bibinfo  {journal} {Phys. Rev.}\ }\textbf {\bibinfo {volume} {56}},\
  \bibinfo {pages} {340--343} (\bibinfo {year} {1939})}\BibitemShut {NoStop}%
\bibitem [{\citenamefont {Singh}\ and\ \citenamefont
  {Singh}(1989)}]{10.1119/1.15842}%
  \BibitemOpen
  \bibfield  {author} {\bibinfo {author} {\bibfnamefont {S.~Brajamani}\
  \bibnamefont {Singh}}\ and\ \bibinfo {author} {\bibfnamefont {C.~A.}\
  \bibnamefont {Singh}},\ }\bibfield  {title} {\enquote {\bibinfo {title}
  {{Extensions of the Feynman–Hellman theorem and applications}},}\ }\href
  {\doibase 10.1119/1.15842} {\bibfield  {journal} {\bibinfo  {journal}
  {American Journal of Physics}\ }\textbf {\bibinfo {volume} {57}},\ \bibinfo
  {pages} {894--899} (\bibinfo {year} {1989})},\ \Eprint
  {http://arxiv.org/abs/https://pubs.aip.org/aapt/ajp/article-pdf/57/10/894/8503482/894\_1\_online.pdf}
  {https://pubs.aip.org/aapt/ajp/article-pdf/57/10/894/8503482/894\_1\_online.pdf}
  \BibitemShut {NoStop}%
\bibitem [{\citenamefont {{G{\"u}ttinger}}(1932)}]{1932ZPhy...73..169G}%
  \BibitemOpen
  \bibfield  {author} {\bibinfo {author} {\bibfnamefont {P.}~\bibnamefont
  {{G{\"u}ttinger}}},\ }\bibfield  {title} {\enquote {\bibinfo {title} {{Das
  Verhalten von Atomen im magnetischen Drehfeld}},}\ }\href {\doibase
  10.1007/BF01351211} {\bibfield  {journal} {\bibinfo  {journal} {Zeitschrift
  fur Physik}\ }\textbf {\bibinfo {volume} {73}},\ \bibinfo {pages} {169--184}
  (\bibinfo {year} {1932})}\BibitemShut {NoStop}%
\bibitem [{\citenamefont {Hellmann}(2015)}]{hellmann1937}%
  \BibitemOpen
  \bibfield  {author} {\bibinfo {author} {\bibfnamefont {Hans}\ \bibnamefont
  {Hellmann}},\ }\href {\doibase https://doi.org/10.1007/978-3-662-45967-6}
  {\emph {\bibinfo {title} {Hans Hellmann: Einführung in die
  Quantenchemie}}},\ edited by\ \bibinfo {editor} {\bibfnamefont {Dirk}\
  \bibnamefont {Andrae}}\ (\bibinfo  {publisher} {Springer Spektrum Berlin,
  Heidelberg},\ \bibinfo {year} {2015})\ pp.\ \bibinfo {pages} {VII,
  389}\BibitemShut {NoStop}%
\bibitem [{\citenamefont {Fitts}(1999)}]{fitts_1999}%
  \BibitemOpen
  \bibfield  {author} {\bibinfo {author} {\bibfnamefont {Donald~D.}\
  \bibnamefont {Fitts}},\ }\href {\doibase 10.1017/CBO9780511813542} {\emph
  {\bibinfo {title} {Principles of Quantum Mechanics: As Applied to Chemistry
  and Chemical Physics}}}\ (\bibinfo  {publisher} {Cambridge University
  Press},\ \bibinfo {year} {1999})\BibitemShut {NoStop}%
\bibitem [{\citenamefont {Fernández}(2022)}]{FERNANDEZ2022169158}%
  \BibitemOpen
  \bibfield  {author} {\bibinfo {author} {\bibfnamefont {Francisco~M.}\
  \bibnamefont {Fernández}},\ }\bibfield  {title} {\enquote {\bibinfo {title}
  {Comment on “pdm klein–gordon oscillators in cosmic string spacetime in
  magnetic and aharonov–bohm flux fields within the kaluza–klein
  theory”},}\ }\href {\doibase https://doi.org/10.1016/j.aop.2022.169158}
  {\bibfield  {journal} {\bibinfo  {journal} {Annals of Physics}\ }\textbf
  {\bibinfo {volume} {447}},\ \bibinfo {pages} {169158} (\bibinfo {year}
  {2022})}\BibitemShut {NoStop}%
\bibitem [{\citenamefont {Brice\~no}\ \emph {et~al.}(2020)\citenamefont
  {Brice\~no}, \citenamefont {Hansen},\ and\ \citenamefont
  {Jackura}}]{PhysRevD.101.094508}%
  \BibitemOpen
  \bibfield  {author} {\bibinfo {author} {\bibfnamefont {Ra\'ul~A.}\
  \bibnamefont {Brice\~no}}, \bibinfo {author} {\bibfnamefont {Maxwell~T.}\
  \bibnamefont {Hansen}}, \ and\ \bibinfo {author} {\bibfnamefont {Andrew~W.}\
  \bibnamefont {Jackura}},\ }\bibfield  {title} {\enquote {\bibinfo {title}
  {Consistency checks for two-body finite-volume matrix elements. ii.
  perturbative systems},}\ }\href {\doibase 10.1103/PhysRevD.101.094508}
  {\bibfield  {journal} {\bibinfo  {journal} {Phys. Rev. D}\ }\textbf {\bibinfo
  {volume} {101}},\ \bibinfo {pages} {094508} (\bibinfo {year}
  {2020})}\BibitemShut {NoStop}%
\bibitem [{\citenamefont {Maslov}\ and\ \citenamefont
  {Blaschke}(2023)}]{PhysRevD.107.094010}%
  \BibitemOpen
  \bibfield  {author} {\bibinfo {author} {\bibfnamefont {K.}~\bibnamefont
  {Maslov}}\ and\ \bibinfo {author} {\bibfnamefont {D.}~\bibnamefont
  {Blaschke}},\ }\bibfield  {title} {\enquote {\bibinfo {title} {Effect of
  mesonic off-shell correlations in the pnjl equation of state},}\ }\href
  {\doibase 10.1103/PhysRevD.107.094010} {\bibfield  {journal} {\bibinfo
  {journal} {Phys. Rev. D}\ }\textbf {\bibinfo {volume} {107}},\ \bibinfo
  {pages} {094010} (\bibinfo {year} {2023})}\BibitemShut {NoStop}%
\bibitem [{\citenamefont {Huang}\ \emph {et~al.}(2020)\citenamefont {Huang},
  \citenamefont {Sun},\ and\ \citenamefont {Chen}}]{PhysRevD.101.054007}%
  \BibitemOpen
  \bibfield  {author} {\bibinfo {author} {\bibfnamefont {Jing-Hui}\
  \bibnamefont {Huang}}, \bibinfo {author} {\bibfnamefont {Ting-Ting}\
  \bibnamefont {Sun}}, \ and\ \bibinfo {author} {\bibfnamefont {Huan}\
  \bibnamefont {Chen}},\ }\bibfield  {title} {\enquote {\bibinfo {title}
  {Evaluation of pion-nucleon sigma term in dyson-schwinger equation approach
  of qcd},}\ }\href {\doibase 10.1103/PhysRevD.101.054007} {\bibfield
  {journal} {\bibinfo  {journal} {Phys. Rev. D}\ }\textbf {\bibinfo {volume}
  {101}},\ \bibinfo {pages} {054007} (\bibinfo {year} {2020})}\BibitemShut
  {NoStop}%
\bibitem [{\citenamefont {Sen}\ \emph {et~al.}(2021)\citenamefont {Sen},
  \citenamefont {Petschlies}, \citenamefont {Schlage},\ and\ \citenamefont
  {Urbach}}]{sen2021hadronic}%
  \BibitemOpen
  \bibfield  {author} {\bibinfo {author} {\bibfnamefont {Aniket}\ \bibnamefont
  {Sen}}, \bibinfo {author} {\bibfnamefont {Marcus}\ \bibnamefont
  {Petschlies}}, \bibinfo {author} {\bibfnamefont {Nikolas}\ \bibnamefont
  {Schlage}}, \ and\ \bibinfo {author} {\bibfnamefont {Carsten}\ \bibnamefont
  {Urbach}},\ }\bibfield  {title} {\enquote {\bibinfo {title} {Hadronic parity
  violation from 4-quark interactions},}\ }\href@noop {} {\  (\bibinfo {year}
  {2021})},\ \Eprint {http://arxiv.org/abs/2111.09025} {arXiv:2111.09025
  [hep-lat]} \BibitemShut {NoStop}%
\bibitem [{\citenamefont {Durr}\ \emph {et~al.}(2016)\citenamefont {Durr},
  \citenamefont {Fodor}, \citenamefont {Hoelbling}, \citenamefont {Katz},
  \citenamefont {Krieg}, \citenamefont {Lellouch}, \citenamefont {Lippert},
  \citenamefont {Metivet}, \citenamefont {Portelli}, \citenamefont {Szabo},
  \citenamefont {Torrero}, \citenamefont {Toth},\ and\ \citenamefont
  {Varnhorst}}]{PhysRevLett.116.172001}%
  \BibitemOpen
  \bibfield  {author} {\bibinfo {author} {\bibfnamefont {S.}~\bibnamefont
  {Durr}}, \bibinfo {author} {\bibfnamefont {Z.}~\bibnamefont {Fodor}},
  \bibinfo {author} {\bibfnamefont {C.}~\bibnamefont {Hoelbling}}, \bibinfo
  {author} {\bibfnamefont {S.~D.}\ \bibnamefont {Katz}}, \bibinfo {author}
  {\bibfnamefont {S.}~\bibnamefont {Krieg}}, \bibinfo {author} {\bibfnamefont
  {L.}~\bibnamefont {Lellouch}}, \bibinfo {author} {\bibfnamefont
  {T.}~\bibnamefont {Lippert}}, \bibinfo {author} {\bibfnamefont
  {T.}~\bibnamefont {Metivet}}, \bibinfo {author} {\bibfnamefont
  {A.}~\bibnamefont {Portelli}}, \bibinfo {author} {\bibfnamefont {K.~K.}\
  \bibnamefont {Szabo}}, \bibinfo {author} {\bibfnamefont {C.}~\bibnamefont
  {Torrero}}, \bibinfo {author} {\bibfnamefont {B.~C.}\ \bibnamefont {Toth}}, \
  and\ \bibinfo {author} {\bibfnamefont {L.}~\bibnamefont {Varnhorst}}
  (\bibinfo {collaboration} {Budapest-Marseille-Wuppertal Collaboration}),\
  }\bibfield  {title} {\enquote {\bibinfo {title} {Lattice computation of the
  nucleon scalar quark contents at the physical point},}\ }\href {\doibase
  10.1103/PhysRevLett.116.172001} {\bibfield  {journal} {\bibinfo  {journal}
  {Phys. Rev. Lett.}\ }\textbf {\bibinfo {volume} {116}},\ \bibinfo {pages}
  {172001} (\bibinfo {year} {2016})}\BibitemShut {NoStop}%
\bibitem [{\citenamefont {Can}\ \emph {et~al.}(2020)\citenamefont {Can},
  \citenamefont {Hannaford-Gunn}, \citenamefont {Horsley}, \citenamefont
  {Nakamura}, \citenamefont {Perlt}, \citenamefont {Rakow}, \citenamefont
  {Schierholz}, \citenamefont {Somfleth}, \citenamefont {St\"uben},
  \citenamefont {Young},\ and\ \citenamefont {Zanotti}}]{PhysRevD.102.114505}%
  \BibitemOpen
  \bibfield  {author} {\bibinfo {author} {\bibfnamefont {K.~U.}\ \bibnamefont
  {Can}}, \bibinfo {author} {\bibfnamefont {A.}~\bibnamefont {Hannaford-Gunn}},
  \bibinfo {author} {\bibfnamefont {R.}~\bibnamefont {Horsley}}, \bibinfo
  {author} {\bibfnamefont {Y.}~\bibnamefont {Nakamura}}, \bibinfo {author}
  {\bibfnamefont {H.}~\bibnamefont {Perlt}}, \bibinfo {author} {\bibfnamefont
  {P.~E.~L.}\ \bibnamefont {Rakow}}, \bibinfo {author} {\bibfnamefont
  {G.}~\bibnamefont {Schierholz}}, \bibinfo {author} {\bibfnamefont {K.~Y.}\
  \bibnamefont {Somfleth}}, \bibinfo {author} {\bibfnamefont {H.}~\bibnamefont
  {St\"uben}}, \bibinfo {author} {\bibfnamefont {R.~D.}\ \bibnamefont {Young}},
  \ and\ \bibinfo {author} {\bibfnamefont {J.~M.}\ \bibnamefont {Zanotti}}
  (\bibinfo {collaboration} {QCDSF/UKQCD/CSSM Collaborations}),\ }\bibfield
  {title} {\enquote {\bibinfo {title} {Lattice qcd evaluation of the compton
  amplitude employing the feynman-hellmann theorem},}\ }\href {\doibase
  10.1103/PhysRevD.102.114505} {\bibfield  {journal} {\bibinfo  {journal}
  {Phys. Rev. D}\ }\textbf {\bibinfo {volume} {102}},\ \bibinfo {pages}
  {114505} (\bibinfo {year} {2020})}\BibitemShut {NoStop}%
\bibitem [{\citenamefont {Batelaan}\ \emph
  {et~al.}(2023{\natexlab{a}})\citenamefont {Batelaan}, \citenamefont {Can},
  \citenamefont {Hannaford-Gunn}, \citenamefont {Horsley}, \citenamefont
  {Nakamura}, \citenamefont {Perlt}, \citenamefont {Rakow}, \citenamefont
  {Schierholz}, \citenamefont {St\"uben}, \citenamefont {Young},\ and\
  \citenamefont {Zanotti}}]{PhysRevD.107.054503}%
  \BibitemOpen
  \bibfield  {author} {\bibinfo {author} {\bibfnamefont {M.}~\bibnamefont
  {Batelaan}}, \bibinfo {author} {\bibfnamefont {K.~U.}\ \bibnamefont {Can}},
  \bibinfo {author} {\bibfnamefont {A.}~\bibnamefont {Hannaford-Gunn}},
  \bibinfo {author} {\bibfnamefont {R.}~\bibnamefont {Horsley}}, \bibinfo
  {author} {\bibfnamefont {Y.}~\bibnamefont {Nakamura}}, \bibinfo {author}
  {\bibfnamefont {H.}~\bibnamefont {Perlt}}, \bibinfo {author} {\bibfnamefont
  {P.~E.~L.}\ \bibnamefont {Rakow}}, \bibinfo {author} {\bibfnamefont
  {G.}~\bibnamefont {Schierholz}}, \bibinfo {author} {\bibfnamefont
  {H.}~\bibnamefont {St\"uben}}, \bibinfo {author} {\bibfnamefont {R.~D.}\
  \bibnamefont {Young}}, \ and\ \bibinfo {author} {\bibfnamefont {J.~M.}\
  \bibnamefont {Zanotti}} (\bibinfo {collaboration} {QCDSF/UKQCD/CSSM
  Collaborations}),\ }\bibfield  {title} {\enquote {\bibinfo {title} {Moments
  and power corrections of longitudinal and transverse proton structure
  functions from lattice qcd},}\ }\href {\doibase 10.1103/PhysRevD.107.054503}
  {\bibfield  {journal} {\bibinfo  {journal} {Phys. Rev. D}\ }\textbf {\bibinfo
  {volume} {107}},\ \bibinfo {pages} {054503} (\bibinfo {year}
  {2023}{\natexlab{a}})}\BibitemShut {NoStop}%
\bibitem [{\citenamefont {Takeda}\ \emph {et~al.}(2011)\citenamefont {Takeda},
  \citenamefont {Aoki}, \citenamefont {Hashimoto}, \citenamefont {Kaneko},
  \citenamefont {Noaki},\ and\ \citenamefont {Onogi}}]{PhysRevD.83.114506}%
  \BibitemOpen
  \bibfield  {author} {\bibinfo {author} {\bibfnamefont {K.}~\bibnamefont
  {Takeda}}, \bibinfo {author} {\bibfnamefont {S.}~\bibnamefont {Aoki}},
  \bibinfo {author} {\bibfnamefont {S.}~\bibnamefont {Hashimoto}}, \bibinfo
  {author} {\bibfnamefont {T.}~\bibnamefont {Kaneko}}, \bibinfo {author}
  {\bibfnamefont {J.}~\bibnamefont {Noaki}}, \ and\ \bibinfo {author}
  {\bibfnamefont {T.}~\bibnamefont {Onogi}} (\bibinfo {collaboration} {JLQCD
  Collaboration}),\ }\bibfield  {title} {\enquote {\bibinfo {title} {Nucleon
  strange quark content from two-flavor lattice qcd with exact chiral
  symmetry},}\ }\href {\doibase 10.1103/PhysRevD.83.114506} {\bibfield
  {journal} {\bibinfo  {journal} {Phys. Rev. D}\ }\textbf {\bibinfo {volume}
  {83}},\ \bibinfo {pages} {114506} (\bibinfo {year} {2011})}\BibitemShut
  {NoStop}%
\bibitem [{\citenamefont {Bouchard}\ \emph {et~al.}(2017)\citenamefont
  {Bouchard}, \citenamefont {Chang}, \citenamefont {Kurth}, \citenamefont
  {Orginos},\ and\ \citenamefont {Walker-Loud}}]{PhysRevD.96.014504}%
  \BibitemOpen
  \bibfield  {author} {\bibinfo {author} {\bibfnamefont {Chris}\ \bibnamefont
  {Bouchard}}, \bibinfo {author} {\bibfnamefont {Chia~Cheng}\ \bibnamefont
  {Chang}}, \bibinfo {author} {\bibfnamefont {Thorsten}\ \bibnamefont {Kurth}},
  \bibinfo {author} {\bibfnamefont {Kostas}\ \bibnamefont {Orginos}}, \ and\
  \bibinfo {author} {\bibfnamefont {Andr\'e}\ \bibnamefont {Walker-Loud}},\
  }\bibfield  {title} {\enquote {\bibinfo {title} {On the feynman-hellmann
  theorem in quantum field theory and the calculation of matrix elements},}\
  }\href {\doibase 10.1103/PhysRevD.96.014504} {\bibfield  {journal} {\bibinfo
  {journal} {Phys. Rev. D}\ }\textbf {\bibinfo {volume} {96}},\ \bibinfo
  {pages} {014504} (\bibinfo {year} {2017})}\BibitemShut {NoStop}%
\bibitem [{\citenamefont {Qin}\ \emph {et~al.}(2022)\citenamefont {Qin},
  \citenamefont {Bai}, \citenamefont {Chen},\ and\ \citenamefont
  {Qin}}]{PhysRevD.106.034006}%
  \BibitemOpen
  \bibfield  {author} {\bibinfo {author} {\bibfnamefont {Pianpian}\
  \bibnamefont {Qin}}, \bibinfo {author} {\bibfnamefont {Zhan}\ \bibnamefont
  {Bai}}, \bibinfo {author} {\bibfnamefont {Muyang}\ \bibnamefont {Chen}}, \
  and\ \bibinfo {author} {\bibfnamefont {Si-xue}\ \bibnamefont {Qin}},\
  }\bibfield  {title} {\enquote {\bibinfo {title} {Partial wave analysis for
  the in-hadron condensate},}\ }\href {\doibase 10.1103/PhysRevD.106.034006}
  {\bibfield  {journal} {\bibinfo  {journal} {Phys. Rev. D}\ }\textbf {\bibinfo
  {volume} {106}},\ \bibinfo {pages} {034006} (\bibinfo {year}
  {2022})}\BibitemShut {NoStop}%
\bibitem [{\citenamefont {Batelaan}\ \emph
  {et~al.}(2023{\natexlab{b}})\citenamefont {Batelaan}, \citenamefont {Can},
  \citenamefont {Horsley}, \citenamefont {Nakamura}, \citenamefont {Perlt},
  \citenamefont {Rakow}, \citenamefont {Schierholz}, \citenamefont {Stüben},
  \citenamefont {Young},\ and\ \citenamefont
  {Zanotti}}]{batelaan2023quasidegenerate}%
  \BibitemOpen
  \bibfield  {author} {\bibinfo {author} {\bibfnamefont {M.}~\bibnamefont
  {Batelaan}}, \bibinfo {author} {\bibfnamefont {K.~U.}\ \bibnamefont {Can}},
  \bibinfo {author} {\bibfnamefont {R.}~\bibnamefont {Horsley}}, \bibinfo
  {author} {\bibfnamefont {Y.}~\bibnamefont {Nakamura}}, \bibinfo {author}
  {\bibfnamefont {H.}~\bibnamefont {Perlt}}, \bibinfo {author} {\bibfnamefont
  {P.~E.~L.}\ \bibnamefont {Rakow}}, \bibinfo {author} {\bibfnamefont
  {G.}~\bibnamefont {Schierholz}}, \bibinfo {author} {\bibfnamefont
  {H.}~\bibnamefont {Stüben}}, \bibinfo {author} {\bibfnamefont {R.~D.}\
  \bibnamefont {Young}}, \ and\ \bibinfo {author} {\bibfnamefont {J.~M.}\
  \bibnamefont {Zanotti}},\ }\bibfield  {title} {\enquote {\bibinfo {title}
  {Quasi-degenerate baryon energy states, the feynman--hellmann theorem and
  transition matrix elements},}\ }\href@noop {} {\  (\bibinfo {year}
  {2023}{\natexlab{b}})},\ \Eprint {http://arxiv.org/abs/2302.04911}
  {arXiv:2302.04911 [hep-lat]} \BibitemShut {NoStop}%
\bibitem [{\citenamefont {Hannaford-Gunn}\ \emph {et~al.}(2020)\citenamefont
  {Hannaford-Gunn}, \citenamefont {Horsley}, \citenamefont {Nakamura},
  \citenamefont {Perlt}, \citenamefont {Rakow}, \citenamefont {Schierholz},
  \citenamefont {Somfleth}, \citenamefont {Stüben}, \citenamefont {Young},\
  and\ \citenamefont {Zanotti}}]{hannafordgunn2020scaling}%
  \BibitemOpen
  \bibfield  {author} {\bibinfo {author} {\bibfnamefont {A.}~\bibnamefont
  {Hannaford-Gunn}}, \bibinfo {author} {\bibfnamefont {R.}~\bibnamefont
  {Horsley}}, \bibinfo {author} {\bibfnamefont {Y.}~\bibnamefont {Nakamura}},
  \bibinfo {author} {\bibfnamefont {H.}~\bibnamefont {Perlt}}, \bibinfo
  {author} {\bibfnamefont {P.~E.~L.}\ \bibnamefont {Rakow}}, \bibinfo {author}
  {\bibfnamefont {G.}~\bibnamefont {Schierholz}}, \bibinfo {author}
  {\bibfnamefont {K.}~\bibnamefont {Somfleth}}, \bibinfo {author}
  {\bibfnamefont {H.}~\bibnamefont {Stüben}}, \bibinfo {author} {\bibfnamefont
  {R.~D.}\ \bibnamefont {Young}}, \ and\ \bibinfo {author} {\bibfnamefont
  {J.~M.}\ \bibnamefont {Zanotti}},\ }\bibfield  {title} {\enquote {\bibinfo
  {title} {Scaling and higher twist in the nucleon compton amplitude},}\
  }\href@noop {} {\  (\bibinfo {year} {2020})},\ \Eprint
  {http://arxiv.org/abs/2001.05090} {arXiv:2001.05090 [hep-lat]} \BibitemShut
  {NoStop}%
\bibitem [{\citenamefont {Lacour}\ \emph {et~al.}(2010)\citenamefont {Lacour},
  \citenamefont {Oller},\ and\ \citenamefont {Meißner}}]{Lacour_2010}%
  \BibitemOpen
  \bibfield  {author} {\bibinfo {author} {\bibfnamefont {A}~\bibnamefont
  {Lacour}}, \bibinfo {author} {\bibfnamefont {J~A}\ \bibnamefont {Oller}}, \
  and\ \bibinfo {author} {\bibfnamefont {U-G}\ \bibnamefont {Meißner}},\
  }\bibfield  {title} {\enquote {\bibinfo {title} {The chiral quark condensate
  and pion decay constant in nuclear matter at next-to-leading order},}\ }\href
  {\doibase 10.1088/0954-3899/37/12/125002} {\bibfield  {journal} {\bibinfo
  {journal} {Journal of Physics G: Nuclear and Particle Physics}\ }\textbf
  {\bibinfo {volume} {37}},\ \bibinfo {pages} {125002} (\bibinfo {year}
  {2010})}\BibitemShut {NoStop}%
\bibitem [{\citenamefont {Can}\ \emph {et~al.}(2022)\citenamefont {Can},
  \citenamefont {Hannaford-Gunn}, \citenamefont {Horsley}, \citenamefont
  {Nakamura}, \citenamefont {Perlt}, \citenamefont {Rakow}, \citenamefont
  {Sankey}, \citenamefont {Schierholz}, \citenamefont {Stüben}, \citenamefont
  {Young},\ and\ \citenamefont {Zanotti}}]{can2022compton}%
  \BibitemOpen
  \bibfield  {author} {\bibinfo {author} {\bibfnamefont {K.~U.}\ \bibnamefont
  {Can}}, \bibinfo {author} {\bibfnamefont {A.}~\bibnamefont {Hannaford-Gunn}},
  \bibinfo {author} {\bibfnamefont {R.}~\bibnamefont {Horsley}}, \bibinfo
  {author} {\bibfnamefont {Y.}~\bibnamefont {Nakamura}}, \bibinfo {author}
  {\bibfnamefont {H.}~\bibnamefont {Perlt}}, \bibinfo {author} {\bibfnamefont
  {P.~E.~L.}\ \bibnamefont {Rakow}}, \bibinfo {author} {\bibfnamefont
  {E.}~\bibnamefont {Sankey}}, \bibinfo {author} {\bibfnamefont
  {G.}~\bibnamefont {Schierholz}}, \bibinfo {author} {\bibfnamefont
  {H.}~\bibnamefont {Stüben}}, \bibinfo {author} {\bibfnamefont {R.~D.}\
  \bibnamefont {Young}}, \ and\ \bibinfo {author} {\bibfnamefont {J.~M.}\
  \bibnamefont {Zanotti}},\ }\bibfield  {title} {\enquote {\bibinfo {title}
  {The compton amplitude, lattice qcd and the feynman-hellmann approach},}\
  }\href@noop {} {\  (\bibinfo {year} {2022})},\ \Eprint
  {http://arxiv.org/abs/2201.08367} {arXiv:2201.08367 [hep-lat]} \BibitemShut
  {NoStop}%
\bibitem [{Phy()}]{PhysRevB.100.024401}%
  \BibitemOpen
  \href@noop {} {\ }\BibitemShut {NoStop}%
\bibitem [{\citenamefont {Krotz}\ \emph {et~al.}(2021)\citenamefont {Krotz},
  \citenamefont {Provazza},\ and\ \citenamefont
  {Tempelaar}}]{10.1063/5.0053177}%
  \BibitemOpen
  \bibfield  {author} {\bibinfo {author} {\bibfnamefont {Alex}\ \bibnamefont
  {Krotz}}, \bibinfo {author} {\bibfnamefont {Justin}\ \bibnamefont
  {Provazza}}, \ and\ \bibinfo {author} {\bibfnamefont {Roel}\ \bibnamefont
  {Tempelaar}},\ }\bibfield  {title} {\enquote {\bibinfo {title} {{A
  reciprocal-space formulation of mixed quantum–classical dynamics}},}\
  }\href {\doibase 10.1063/5.0053177} {\bibfield  {journal} {\bibinfo
  {journal} {The Journal of Chemical Physics}\ }\textbf {\bibinfo {volume}
  {154}} (\bibinfo {year} {2021}),\ 10.1063/5.0053177},\ \bibinfo {note}
  {224101},\ \Eprint
  {http://arxiv.org/abs/https://pubs.aip.org/aip/jcp/article-pdf/doi/10.1063/5.0053177/14002402/224101\_1\_online.pdf}
  {https://pubs.aip.org/aip/jcp/article-pdf/doi/10.1063/5.0053177/14002402/224101\_1\_online.pdf}
  \BibitemShut {NoStop}%
\bibitem [{\citenamefont {Chen}\ and\ \citenamefont
  {Zhang}(2023)}]{PhysRevB.107.195150}%
  \BibitemOpen
  \bibfield  {author} {\bibinfo {author} {\bibfnamefont {Siyuan}\ \bibnamefont
  {Chen}}\ and\ \bibinfo {author} {\bibfnamefont {Shiwei}\ \bibnamefont
  {Zhang}},\ }\bibfield  {title} {\enquote {\bibinfo {title} {Computation of
  forces and stresses in solids: Towards accurate structural optimization with
  auxiliary-field quantum monte carlo},}\ }\href {\doibase
  10.1103/PhysRevB.107.195150} {\bibfield  {journal} {\bibinfo  {journal}
  {Phys. Rev. B}\ }\textbf {\bibinfo {volume} {107}},\ \bibinfo {pages}
  {195150} (\bibinfo {year} {2023})}\BibitemShut {NoStop}%
\bibitem [{\citenamefont {Rosi}\ \emph {et~al.}(2023)\citenamefont {Rosi},
  \citenamefont {Rota}, \citenamefont {Astrakharchik},\ and\ \citenamefont
  {Boronat}}]{De_Rosi_2023}%
  \BibitemOpen
  \bibfield  {author} {\bibinfo {author} {\bibfnamefont {Giulia~De}\
  \bibnamefont {Rosi}}, \bibinfo {author} {\bibfnamefont {Riccardo}\
  \bibnamefont {Rota}}, \bibinfo {author} {\bibfnamefont {Grigori~E}\
  \bibnamefont {Astrakharchik}}, \ and\ \bibinfo {author} {\bibfnamefont
  {Jordi}\ \bibnamefont {Boronat}},\ }\bibfield  {title} {\enquote {\bibinfo
  {title} {Correlation properties of a one-dimensional repulsive bose gas at
  finite temperature},}\ }\href {\doibase 10.1088/1367-2630/acc6e6} {\bibfield
  {journal} {\bibinfo  {journal} {New Journal of Physics}\ }\textbf {\bibinfo
  {volume} {25}},\ \bibinfo {pages} {043002} (\bibinfo {year}
  {2023})}\BibitemShut {NoStop}%
\bibitem [{\citenamefont {Rufus}\ and\ \citenamefont
  {Gavini}(2022)}]{PhysRevB.106.085108}%
  \BibitemOpen
  \bibfield  {author} {\bibinfo {author} {\bibfnamefont {Nelson~D.}\
  \bibnamefont {Rufus}}\ and\ \bibinfo {author} {\bibfnamefont {Vikram}\
  \bibnamefont {Gavini}},\ }\bibfield  {title} {\enquote {\bibinfo {title}
  {Ionic forces and stress tensor in all-electron density functional theory
  calculations using an enriched finite-element basis},}\ }\href {\doibase
  10.1103/PhysRevB.106.085108} {\bibfield  {journal} {\bibinfo  {journal}
  {Phys. Rev. B}\ }\textbf {\bibinfo {volume} {106}},\ \bibinfo {pages}
  {085108} (\bibinfo {year} {2022})}\BibitemShut {NoStop}%
\bibitem [{\citenamefont {Zwanziger}\ \emph {et~al.}(2023)\citenamefont
  {Zwanziger}, \citenamefont {Torrent},\ and\ \citenamefont
  {Gonze}}]{PhysRevB.107.165157}%
  \BibitemOpen
  \bibfield  {author} {\bibinfo {author} {\bibfnamefont {J.~W.}\ \bibnamefont
  {Zwanziger}}, \bibinfo {author} {\bibfnamefont {M.}~\bibnamefont {Torrent}},
  \ and\ \bibinfo {author} {\bibfnamefont {X.}~\bibnamefont {Gonze}},\
  }\bibfield  {title} {\enquote {\bibinfo {title} {Orbital magnetism and
  chemical shielding in the projector augmented-wave formalism},}\ }\href
  {\doibase 10.1103/PhysRevB.107.165157} {\bibfield  {journal} {\bibinfo
  {journal} {Phys. Rev. B}\ }\textbf {\bibinfo {volume} {107}},\ \bibinfo
  {pages} {165157} (\bibinfo {year} {2023})}\BibitemShut {NoStop}%
\bibitem [{\citenamefont {Pakizer}\ and\ \citenamefont
  {Matos-Abiague}(2021)}]{PhysRevB.104.L100506}%
  \BibitemOpen
  \bibfield  {author} {\bibinfo {author} {\bibfnamefont {Joseph~D.}\
  \bibnamefont {Pakizer}}\ and\ \bibinfo {author} {\bibfnamefont {Alex}\
  \bibnamefont {Matos-Abiague}},\ }\bibfield  {title} {\enquote {\bibinfo
  {title} {Signatures of topological transitions in the spin susceptibility of
  josephson junctions},}\ }\href {\doibase 10.1103/PhysRevB.104.L100506}
  {\bibfield  {journal} {\bibinfo  {journal} {Phys. Rev. B}\ }\textbf {\bibinfo
  {volume} {104}},\ \bibinfo {pages} {L100506} (\bibinfo {year}
  {2021})}\BibitemShut {NoStop}%
\bibitem [{\citenamefont {Etea}\ and\ \citenamefont
  {Nigussa}(2023)}]{etea2023study}%
  \BibitemOpen
  \bibfield  {author} {\bibinfo {author} {\bibfnamefont {H.~D.}\ \bibnamefont
  {Etea}}\ and\ \bibinfo {author} {\bibfnamefont {K.~N.}\ \bibnamefont
  {Nigussa}},\ }\bibfield  {title} {\enquote {\bibinfo {title} {Study of novel
  properties of graphene-zno heterojunction interface using density functional
  theory},}\ }\href@noop {} {\  (\bibinfo {year} {2023})},\ \Eprint
  {http://arxiv.org/abs/2305.02798} {arXiv:2305.02798 [cond-mat.mtrl-sci]}
  \BibitemShut {NoStop}%
\bibitem [{\citenamefont {Karaca}\ and\ \citenamefont
  {Temizer}(2023)}]{KARACA2023115674}%
  \BibitemOpen
  \bibfield  {author} {\bibinfo {author} {\bibfnamefont {K.}~\bibnamefont
  {Karaca}}\ and\ \bibinfo {author} {\bibfnamefont {İ.}\ \bibnamefont
  {Temizer}},\ }\bibfield  {title} {\enquote {\bibinfo {title} {Variationally
  consistent hellmann–feynman forces in the finite element formulation of
  kohn–sham density functional theory},}\ }\href {\doibase
  https://doi.org/10.1016/j.cma.2022.115674} {\bibfield  {journal} {\bibinfo
  {journal} {Computer Methods in Applied Mechanics and Engineering}\ }\textbf
  {\bibinfo {volume} {403}},\ \bibinfo {pages} {115674} (\bibinfo {year}
  {2023})}\BibitemShut {NoStop}%
\bibitem [{\citenamefont {Unke}\ \emph {et~al.}(2021)\citenamefont {Unke},
  \citenamefont {Chmiela}, \citenamefont {Sauceda}, \citenamefont {Gastegger},
  \citenamefont {Poltavsky}, \citenamefont {Schütt}, \citenamefont
  {Tkatchenko},\ and\ \citenamefont
  {Müller}}]{doi:10.1021/acs.chemrev.0c01111}%
  \BibitemOpen
  \bibfield  {author} {\bibinfo {author} {\bibfnamefont {Oliver~T.}\
  \bibnamefont {Unke}}, \bibinfo {author} {\bibfnamefont {Stefan}\ \bibnamefont
  {Chmiela}}, \bibinfo {author} {\bibfnamefont {Huziel~E.}\ \bibnamefont
  {Sauceda}}, \bibinfo {author} {\bibfnamefont {Michael}\ \bibnamefont
  {Gastegger}}, \bibinfo {author} {\bibfnamefont {Igor}\ \bibnamefont
  {Poltavsky}}, \bibinfo {author} {\bibfnamefont {Kristof~T.}\ \bibnamefont
  {Schütt}}, \bibinfo {author} {\bibfnamefont {Alexandre}\ \bibnamefont
  {Tkatchenko}}, \ and\ \bibinfo {author} {\bibfnamefont {Klaus-Robert}\
  \bibnamefont {Müller}},\ }\bibfield  {title} {\enquote {\bibinfo {title}
  {Machine learning force fields},}\ }\href {\doibase
  10.1021/acs.chemrev.0c01111} {\bibfield  {journal} {\bibinfo  {journal}
  {Chemical Reviews}\ }\textbf {\bibinfo {volume} {121}},\ \bibinfo {pages}
  {10142--10186} (\bibinfo {year} {2021})},\ \bibinfo {note} {pMID: 33705118},\
  \Eprint {http://arxiv.org/abs/https://doi.org/10.1021/acs.chemrev.0c01111}
  {https://doi.org/10.1021/acs.chemrev.0c01111} \BibitemShut {NoStop}%
\bibitem [{\citenamefont {Liu}(2020)}]{liu2020pseudomass}%
  \BibitemOpen
  \bibfield  {author} {\bibinfo {author} {\bibfnamefont {Qing-Long}\
  \bibnamefont {Liu}},\ }\bibfield  {title} {\enquote {\bibinfo {title}
  {Pseudo-mass parameterized alchemical equation: a generalisation of the
  molecular schr\"odinger equation},}\ }\href@noop {} {\  (\bibinfo {year}
  {2020})},\ \Eprint {http://arxiv.org/abs/2012.00843} {arXiv:2012.00843
  [physics.chem-ph]} \BibitemShut {NoStop}%
\bibitem [{\citenamefont {Fernández}(2019)}]{fernandezdegen}%
  \BibitemOpen
  \bibfield  {author} {\bibinfo {author} {\bibfnamefont {Francisco~M.}\
  \bibnamefont {Fernández}},\ }\bibfield  {title} {\enquote {\bibinfo {title}
  {On the hellmann-feynman theorem for degenerate states},}\ }\href@noop {} {\
  (\bibinfo {year} {2019})},\ \Eprint {http://arxiv.org/abs/1912.04876}
  {arXiv:1912.04876 [quant-ph]} \BibitemShut {NoStop}%
\bibitem [{\citenamefont {Konotop}\ \emph
  {et~al.}(2016{\natexlab{a}})\citenamefont {Konotop}, \citenamefont {Yang},\
  and\ \citenamefont {Zezyulin}}]{rmp_nh}%
  \BibitemOpen
  \bibfield  {author} {\bibinfo {author} {\bibfnamefont {Vladimir~V.}\
  \bibnamefont {Konotop}}, \bibinfo {author} {\bibfnamefont {Jianke}\
  \bibnamefont {Yang}}, \ and\ \bibinfo {author} {\bibfnamefont {Dmitry~A.}\
  \bibnamefont {Zezyulin}},\ }\bibfield  {title} {\enquote {\bibinfo {title}
  {Nonlinear waves in pt-symmetric systems},}\ }\href {\doibase
  10.1103/RevModPhys.88.035002} {\bibfield  {journal} {\bibinfo  {journal}
  {Rev. Mod. Phys.}\ }\textbf {\bibinfo {volume} {88}},\ \bibinfo {pages}
  {035002} (\bibinfo {year} {2016}{\natexlab{a}})}\BibitemShut {NoStop}%
\bibitem [{\citenamefont {Bergholtz}\ \emph {et~al.}(2021)\citenamefont
  {Bergholtz}, \citenamefont {Budich},\ and\ \citenamefont
  {Kunst}}]{rmp_topology}%
  \BibitemOpen
  \bibfield  {author} {\bibinfo {author} {\bibfnamefont {Emil~J.}\ \bibnamefont
  {Bergholtz}}, \bibinfo {author} {\bibfnamefont {Jan~Carl}\ \bibnamefont
  {Budich}}, \ and\ \bibinfo {author} {\bibfnamefont {Flore~K.}\ \bibnamefont
  {Kunst}},\ }\bibfield  {title} {\enquote {\bibinfo {title} {Exceptional
  topology of non-hermitian systems},}\ }\href {\doibase
  10.1103/RevModPhys.93.015005} {\bibfield  {journal} {\bibinfo  {journal}
  {Rev. Mod. Phys.}\ }\textbf {\bibinfo {volume} {93}},\ \bibinfo {pages}
  {015005} (\bibinfo {year} {2021})}\BibitemShut {NoStop}%
\bibitem [{\citenamefont {Makris}\ \emph {et~al.}(2008)\citenamefont {Makris},
  \citenamefont {El-Ganainy}, \citenamefont {Christodoulides},\ and\
  \citenamefont {Musslimani}}]{nh_1}%
  \BibitemOpen
  \bibfield  {author} {\bibinfo {author} {\bibfnamefont {K.~G.}\ \bibnamefont
  {Makris}}, \bibinfo {author} {\bibfnamefont {R.}~\bibnamefont {El-Ganainy}},
  \bibinfo {author} {\bibfnamefont {D.~N.}\ \bibnamefont {Christodoulides}}, \
  and\ \bibinfo {author} {\bibfnamefont {Z.~H.}\ \bibnamefont {Musslimani}},\
  }\bibfield  {title} {\enquote {\bibinfo {title} {Beam dynamics in pt
  symmetric optical lattices},}\ }\href {\doibase
  10.1103/PhysRevLett.100.103904} {\bibfield  {journal} {\bibinfo  {journal}
  {Phys. Rev. Lett.}\ }\textbf {\bibinfo {volume} {100}},\ \bibinfo {pages}
  {103904} (\bibinfo {year} {2008})}\BibitemShut {NoStop}%
\bibitem [{\citenamefont {Klaiman}\ \emph {et~al.}(2008)\citenamefont
  {Klaiman}, \citenamefont {G\"unther},\ and\ \citenamefont {Moiseyev}}]{nh_2}%
  \BibitemOpen
  \bibfield  {author} {\bibinfo {author} {\bibfnamefont {Shachar}\ \bibnamefont
  {Klaiman}}, \bibinfo {author} {\bibfnamefont {Uwe}\ \bibnamefont
  {G\"unther}}, \ and\ \bibinfo {author} {\bibfnamefont {Nimrod}\ \bibnamefont
  {Moiseyev}},\ }\bibfield  {title} {\enquote {\bibinfo {title} {Visualization
  of branch points in pt-symmetric waveguides},}\ }\href {\doibase
  10.1103/PhysRevLett.101.080402} {\bibfield  {journal} {\bibinfo  {journal}
  {Phys. Rev. Lett.}\ }\textbf {\bibinfo {volume} {101}},\ \bibinfo {pages}
  {080402} (\bibinfo {year} {2008})}\BibitemShut {NoStop}%
\bibitem [{\citenamefont {Lin}\ \emph {et~al.}(2011)\citenamefont {Lin},
  \citenamefont {Ramezani}, \citenamefont {Eichelkraut}, \citenamefont
  {Kottos}, \citenamefont {Cao},\ and\ \citenamefont {Christodoulides}}]{nh_3}%
  \BibitemOpen
  \bibfield  {author} {\bibinfo {author} {\bibfnamefont {Zin}\ \bibnamefont
  {Lin}}, \bibinfo {author} {\bibfnamefont {Hamidreza}\ \bibnamefont
  {Ramezani}}, \bibinfo {author} {\bibfnamefont {Toni}\ \bibnamefont
  {Eichelkraut}}, \bibinfo {author} {\bibfnamefont {Tsampikos}\ \bibnamefont
  {Kottos}}, \bibinfo {author} {\bibfnamefont {Hui}\ \bibnamefont {Cao}}, \
  and\ \bibinfo {author} {\bibfnamefont {Demetrios~N.}\ \bibnamefont
  {Christodoulides}},\ }\bibfield  {title} {\enquote {\bibinfo {title}
  {Unidirectional invisibility induced by pt-symmetric periodic structures},}\
  }\href {\doibase 10.1103/PhysRevLett.106.213901} {\bibfield  {journal}
  {\bibinfo  {journal} {Phys. Rev. Lett.}\ }\textbf {\bibinfo {volume} {106}},\
  \bibinfo {pages} {213901} (\bibinfo {year} {2011})}\BibitemShut {NoStop}%
\bibitem [{\citenamefont {Wiersig}(2014)}]{nh_4}%
  \BibitemOpen
  \bibfield  {author} {\bibinfo {author} {\bibfnamefont {Jan}\ \bibnamefont
  {Wiersig}},\ }\bibfield  {title} {\enquote {\bibinfo {title} {Enhancing the
  sensitivity of frequency and energy splitting detection by using exceptional
  points: Application to microcavity sensors for single-particle detection},}\
  }\href {\doibase 10.1103/PhysRevLett.112.203901} {\bibfield  {journal}
  {\bibinfo  {journal} {Phys. Rev. Lett.}\ }\textbf {\bibinfo {volume} {112}},\
  \bibinfo {pages} {203901} (\bibinfo {year} {2014})}\BibitemShut {NoStop}%
\bibitem [{\citenamefont {Hamazaki}\ \emph {et~al.}(2019)\citenamefont
  {Hamazaki}, \citenamefont {Kawabata},\ and\ \citenamefont {Ueda}}]{mbl_1}%
  \BibitemOpen
  \bibfield  {author} {\bibinfo {author} {\bibfnamefont {Ryusuke}\ \bibnamefont
  {Hamazaki}}, \bibinfo {author} {\bibfnamefont {Kohei}\ \bibnamefont
  {Kawabata}}, \ and\ \bibinfo {author} {\bibfnamefont {Masahito}\ \bibnamefont
  {Ueda}},\ }\bibfield  {title} {\enquote {\bibinfo {title} {Non-hermitian
  many-body localization},}\ }\href {\doibase 10.1103/PhysRevLett.123.090603}
  {\bibfield  {journal} {\bibinfo  {journal} {Phys. Rev. Lett.}\ }\textbf
  {\bibinfo {volume} {123}},\ \bibinfo {pages} {090603} (\bibinfo {year}
  {2019})}\BibitemShut {NoStop}%
\bibitem [{\citenamefont {Kawabata}\ and\ \citenamefont {Ryu}(2021)}]{mbl_2}%
  \BibitemOpen
  \bibfield  {author} {\bibinfo {author} {\bibfnamefont {Kohei}\ \bibnamefont
  {Kawabata}}\ and\ \bibinfo {author} {\bibfnamefont {Shinsei}\ \bibnamefont
  {Ryu}},\ }\bibfield  {title} {\enquote {\bibinfo {title} {Nonunitary scaling
  theory of non-hermitian localization},}\ }\href {\doibase
  10.1103/PhysRevLett.126.166801} {\bibfield  {journal} {\bibinfo  {journal}
  {Phys. Rev. Lett.}\ }\textbf {\bibinfo {volume} {126}},\ \bibinfo {pages}
  {166801} (\bibinfo {year} {2021})}\BibitemShut {NoStop}%
\bibitem [{\citenamefont {Pal}\ \emph {et~al.}(2022)\citenamefont {Pal},
  \citenamefont {Modak},\ and\ \citenamefont {Mandal}}]{pal2022dna}%
  \BibitemOpen
  \bibfield  {author} {\bibinfo {author} {\bibfnamefont {Tanmoy}\ \bibnamefont
  {Pal}}, \bibinfo {author} {\bibfnamefont {Ranjan}\ \bibnamefont {Modak}}, \
  and\ \bibinfo {author} {\bibfnamefont {Bhabani~Prasad}\ \bibnamefont
  {Mandal}},\ }\bibfield  {title} {\enquote {\bibinfo {title} {Dna unzipping as
  pt-symmetry breaking transition},}\ }\href@noop {} {\bibfield  {journal}
  {\bibinfo  {journal} {arXiv preprint arXiv:2212.14394}\ } (\bibinfo {year}
  {2022})}\BibitemShut {NoStop}%
\bibitem [{\citenamefont {Rudner}\ and\ \citenamefont {Levitov}(2009)}]{top_1}%
  \BibitemOpen
  \bibfield  {author} {\bibinfo {author} {\bibfnamefont {M.~S.}\ \bibnamefont
  {Rudner}}\ and\ \bibinfo {author} {\bibfnamefont {L.~S.}\ \bibnamefont
  {Levitov}},\ }\bibfield  {title} {\enquote {\bibinfo {title} {Topological
  transition in a non-hermitian quantum walk},}\ }\href {\doibase
  10.1103/PhysRevLett.102.065703} {\bibfield  {journal} {\bibinfo  {journal}
  {Phys. Rev. Lett.}\ }\textbf {\bibinfo {volume} {102}},\ \bibinfo {pages}
  {065703} (\bibinfo {year} {2009})}\BibitemShut {NoStop}%
\bibitem [{\citenamefont {Zeuner}\ \emph {et~al.}(2015)\citenamefont {Zeuner},
  \citenamefont {Rechtsman}, \citenamefont {Plotnik}, \citenamefont {Lumer},
  \citenamefont {Nolte}, \citenamefont {Rudner}, \citenamefont {Segev},\ and\
  \citenamefont {Szameit}}]{top_2}%
  \BibitemOpen
  \bibfield  {author} {\bibinfo {author} {\bibfnamefont {Julia~M.}\
  \bibnamefont {Zeuner}}, \bibinfo {author} {\bibfnamefont {Mikael~C.}\
  \bibnamefont {Rechtsman}}, \bibinfo {author} {\bibfnamefont {Yonatan}\
  \bibnamefont {Plotnik}}, \bibinfo {author} {\bibfnamefont {Yaakov}\
  \bibnamefont {Lumer}}, \bibinfo {author} {\bibfnamefont {Stefan}\
  \bibnamefont {Nolte}}, \bibinfo {author} {\bibfnamefont {Mark~S.}\
  \bibnamefont {Rudner}}, \bibinfo {author} {\bibfnamefont {Mordechai}\
  \bibnamefont {Segev}}, \ and\ \bibinfo {author} {\bibfnamefont {Alexander}\
  \bibnamefont {Szameit}},\ }\bibfield  {title} {\enquote {\bibinfo {title}
  {Observation of a topological transition in the bulk of a non-hermitian
  system},}\ }\href {\doibase 10.1103/PhysRevLett.115.040402} {\bibfield
  {journal} {\bibinfo  {journal} {Phys. Rev. Lett.}\ }\textbf {\bibinfo
  {volume} {115}},\ \bibinfo {pages} {040402} (\bibinfo {year}
  {2015})}\BibitemShut {NoStop}%
\bibitem [{\citenamefont {Lee}(2016)}]{top_3}%
  \BibitemOpen
  \bibfield  {author} {\bibinfo {author} {\bibfnamefont {Tony~E.}\ \bibnamefont
  {Lee}},\ }\bibfield  {title} {\enquote {\bibinfo {title} {Anomalous edge
  state in a non-hermitian lattice},}\ }\href {\doibase
  10.1103/PhysRevLett.116.133903} {\bibfield  {journal} {\bibinfo  {journal}
  {Phys. Rev. Lett.}\ }\textbf {\bibinfo {volume} {116}},\ \bibinfo {pages}
  {133903} (\bibinfo {year} {2016})}\BibitemShut {NoStop}%
\bibitem [{\citenamefont {Xu}\ \emph {et~al.}(2017)\citenamefont {Xu},
  \citenamefont {Wang},\ and\ \citenamefont {Duan}}]{top_4}%
  \BibitemOpen
  \bibfield  {author} {\bibinfo {author} {\bibfnamefont {Yong}\ \bibnamefont
  {Xu}}, \bibinfo {author} {\bibfnamefont {Sheng-Tao}\ \bibnamefont {Wang}}, \
  and\ \bibinfo {author} {\bibfnamefont {L.-M.}\ \bibnamefont {Duan}},\
  }\bibfield  {title} {\enquote {\bibinfo {title} {Weyl exceptional rings in a
  three-dimensional dissipative cold atomic gas},}\ }\href {\doibase
  10.1103/PhysRevLett.118.045701} {\bibfield  {journal} {\bibinfo  {journal}
  {Phys. Rev. Lett.}\ }\textbf {\bibinfo {volume} {118}},\ \bibinfo {pages}
  {045701} (\bibinfo {year} {2017})}\BibitemShut {NoStop}%
\bibitem [{\citenamefont {Dhar}\ \emph {et~al.}(2015)\citenamefont {Dhar},
  \citenamefont {Dasgupta}, \citenamefont {Dhar},\ and\ \citenamefont
  {Sen}}]{dhar_15}%
  \BibitemOpen
  \bibfield  {author} {\bibinfo {author} {\bibfnamefont {Shrabanti}\
  \bibnamefont {Dhar}}, \bibinfo {author} {\bibfnamefont {Subinay}\
  \bibnamefont {Dasgupta}}, \bibinfo {author} {\bibfnamefont {Abhishek}\
  \bibnamefont {Dhar}}, \ and\ \bibinfo {author} {\bibfnamefont {Diptiman}\
  \bibnamefont {Sen}},\ }\bibfield  {title} {\enquote {\bibinfo {title}
  {Detection of a quantum particle on a lattice under repeated projective
  measurements},}\ }\href {\doibase 10.1103/PhysRevA.91.062115} {\bibfield
  {journal} {\bibinfo  {journal} {Phys. Rev. A}\ }\textbf {\bibinfo {volume}
  {91}},\ \bibinfo {pages} {062115} (\bibinfo {year} {2015})}\BibitemShut
  {NoStop}%
\bibitem [{\citenamefont {Modak}\ and\ \citenamefont
  {Aravinda}(2023)}]{modak2023non}%
  \BibitemOpen
  \bibfield  {author} {\bibinfo {author} {\bibfnamefont {Ranjan}\ \bibnamefont
  {Modak}}\ and\ \bibinfo {author} {\bibfnamefont {S}~\bibnamefont
  {Aravinda}},\ }\bibfield  {title} {\enquote {\bibinfo {title} {Non-hermitian
  description of sharp quantum resetting},}\ }\href@noop {} {\bibfield
  {journal} {\bibinfo  {journal} {arXiv preprint arXiv:2303.03790}\ } (\bibinfo
  {year} {2023})}\BibitemShut {NoStop}%
\bibitem [{\citenamefont {Bender}\ \emph {et~al.}(2002)\citenamefont {Bender},
  \citenamefont {Brody},\ and\ \citenamefont {Jones}}]{Bender_2002}%
  \BibitemOpen
  \bibfield  {author} {\bibinfo {author} {\bibfnamefont {Carl~M.}\ \bibnamefont
  {Bender}}, \bibinfo {author} {\bibfnamefont {Dorje~C.}\ \bibnamefont
  {Brody}}, \ and\ \bibinfo {author} {\bibfnamefont {Hugh~F.}\ \bibnamefont
  {Jones}},\ }\bibfield  {title} {\enquote {\bibinfo {title} {Complex extension
  of quantum mechanics},}\ }\href {\doibase 10.1103/PhysRevLett.89.270401}
  {\bibfield  {journal} {\bibinfo  {journal} {Phys. Rev. Lett.}\ }\textbf
  {\bibinfo {volume} {89}},\ \bibinfo {pages} {270401} (\bibinfo {year}
  {2002})}\BibitemShut {NoStop}%
\bibitem [{\citenamefont {Bender}(2005)}]{doi:10.1080/00107500072632}%
  \BibitemOpen
  \bibfield  {author} {\bibinfo {author} {\bibfnamefont {Carl~M}\ \bibnamefont
  {Bender}},\ }\bibfield  {title} {\enquote {\bibinfo {title} {Introduction to
  pt-symmetric quantum theory},}\ }\href {\doibase 10.1080/00107500072632}
  {\bibfield  {journal} {\bibinfo  {journal} {Contemporary Physics}\ }\textbf
  {\bibinfo {volume} {46}},\ \bibinfo {pages} {277--292} (\bibinfo {year}
  {2005})},\ \Eprint
  {http://arxiv.org/abs/https://doi.org/10.1080/00107500072632}
  {https://doi.org/10.1080/00107500072632} \BibitemShut {NoStop}%
\bibitem [{\citenamefont {Khare}\ and\ \citenamefont
  {Mandal}(2000)}]{KHARE200053}%
  \BibitemOpen
  \bibfield  {author} {\bibinfo {author} {\bibfnamefont {Avinash}\ \bibnamefont
  {Khare}}\ and\ \bibinfo {author} {\bibfnamefont {Bhabani~Prasad}\
  \bibnamefont {Mandal}},\ }\bibfield  {title} {\enquote {\bibinfo {title} {A
  pt-invariant potential with complex qes eigenvalues},}\ }\href {\doibase
  https://doi.org/10.1016/S0375-9601(00)00409-6} {\bibfield  {journal}
  {\bibinfo  {journal} {Physics Letters A}\ }\textbf {\bibinfo {volume}
  {272}},\ \bibinfo {pages} {53--56} (\bibinfo {year} {2000})}\BibitemShut
  {NoStop}%
\bibitem [{\citenamefont {Modak}\ and\ \citenamefont
  {Mandal}(2021)}]{PhysRevA.103.062416}%
  \BibitemOpen
  \bibfield  {author} {\bibinfo {author} {\bibfnamefont {Ranjan}\ \bibnamefont
  {Modak}}\ and\ \bibinfo {author} {\bibfnamefont {Bhabani~Prasad}\
  \bibnamefont {Mandal}},\ }\bibfield  {title} {\enquote {\bibinfo {title}
  {Eigenstate entanglement entropy in a pt-invariant non-hermitian system},}\
  }\href {\doibase 10.1103/PhysRevA.103.062416} {\bibfield  {journal} {\bibinfo
   {journal} {Phys. Rev. A}\ }\textbf {\bibinfo {volume} {103}},\ \bibinfo
  {pages} {062416} (\bibinfo {year} {2021})}\BibitemShut {NoStop}%
\bibitem [{\citenamefont {Raval}\ and\ \citenamefont
  {Mandal}(2019)}]{RAVAL2019114699}%
  \BibitemOpen
  \bibfield  {author} {\bibinfo {author} {\bibfnamefont {Haresh}\ \bibnamefont
  {Raval}}\ and\ \bibinfo {author} {\bibfnamefont {Bhabani~Prasad}\
  \bibnamefont {Mandal}},\ }\bibfield  {title} {\enquote {\bibinfo {title}
  {Deconfinement to confinement as pt phase transition},}\ }\href {\doibase
  https://doi.org/10.1016/j.nuclphysb.2019.114699} {\bibfield  {journal}
  {\bibinfo  {journal} {Nuclear Physics B}\ }\textbf {\bibinfo {volume}
  {946}},\ \bibinfo {pages} {114699} (\bibinfo {year} {2019})}\BibitemShut
  {NoStop}%
\bibitem [{\citenamefont {Mandal}(2005)}]{pseudo1}%
  \BibitemOpen
  \bibfield  {author} {\bibinfo {author} {\bibfnamefont {Bhabani~Prasad}\
  \bibnamefont {Mandal}},\ }\bibfield  {title} {\enquote {\bibinfo {title}
  {Pseudo-hermitian interaction between an oscillator and a spin-1/2 particle
  in the external magnetic field},}\ }\href {\doibase
  https://doi.org/10.1142/S0217732305016488} {\bibfield  {journal} {\bibinfo
  {journal} {Modern Physics Letters A}\ }\textbf {\bibinfo {volume} {20}}
  (\bibinfo {year} {2005}),\
  https://doi.org/10.1142/S0217732305016488}\BibitemShut {NoStop}%
\bibitem [{\citenamefont {Mandal}\ \emph {et~al.}(2015)\citenamefont {Mandal},
  \citenamefont {Mourya}, \citenamefont {Ali},\ and\ \citenamefont
  {Ghatak}}]{MANDAL2015185}%
  \BibitemOpen
  \bibfield  {author} {\bibinfo {author} {\bibfnamefont {B.P.}\ \bibnamefont
  {Mandal}}, \bibinfo {author} {\bibfnamefont {B.K.}\ \bibnamefont {Mourya}},
  \bibinfo {author} {\bibfnamefont {K.}~\bibnamefont {Ali}}, \ and\ \bibinfo
  {author} {\bibfnamefont {A.}~\bibnamefont {Ghatak}},\ }\bibfield  {title}
  {\enquote {\bibinfo {title} {Pt phase transition in a (2+1)-d relativistic
  system},}\ }\href {\doibase https://doi.org/10.1016/j.aop.2015.09.022}
  {\bibfield  {journal} {\bibinfo  {journal} {Annals of Physics}\ }\textbf
  {\bibinfo {volume} {363}},\ \bibinfo {pages} {185--193} (\bibinfo {year}
  {2015})}\BibitemShut {NoStop}%
\bibitem [{\citenamefont {Mandal}\ \emph {et~al.}(2013)\citenamefont {Mandal},
  \citenamefont {Mourya},\ and\ \citenamefont {Yadav}}]{MANDAL20131043}%
  \BibitemOpen
  \bibfield  {author} {\bibinfo {author} {\bibfnamefont {Bhabani~Prasad}\
  \bibnamefont {Mandal}}, \bibinfo {author} {\bibfnamefont {Brijesh~Kumar}\
  \bibnamefont {Mourya}}, \ and\ \bibinfo {author} {\bibfnamefont
  {Rajesh~Kumar}\ \bibnamefont {Yadav}},\ }\bibfield  {title} {\enquote
  {\bibinfo {title} {Pt phase transition in higher-dimensional quantum
  systems},}\ }\href {\doibase https://doi.org/10.1016/j.physleta.2013.02.023}
  {\bibfield  {journal} {\bibinfo  {journal} {Physics Letters A}\ }\textbf
  {\bibinfo {volume} {377}},\ \bibinfo {pages} {1043--1046} (\bibinfo {year}
  {2013})}\BibitemShut {NoStop}%
\bibitem [{\citenamefont {Ju}\ \emph {et~al.}(2019{\natexlab{a}})\citenamefont
  {Ju}, \citenamefont {Miranowicz}, \citenamefont {Chen},\ and\ \citenamefont
  {Nori}}]{nogo}%
  \BibitemOpen
  \bibfield  {author} {\bibinfo {author} {\bibfnamefont {Chia-Yi}\ \bibnamefont
  {Ju}}, \bibinfo {author} {\bibfnamefont {Adam}\ \bibnamefont {Miranowicz}},
  \bibinfo {author} {\bibfnamefont {Guang-Yin}\ \bibnamefont {Chen}}, \ and\
  \bibinfo {author} {\bibfnamefont {Franco}\ \bibnamefont {Nori}},\ }\bibfield
  {title} {\enquote {\bibinfo {title} {Non-hermitian hamiltonians and no-go
  theorems in quantum information},}\ }\href {\doibase
  10.1103/PhysRevA.100.062118} {\bibfield  {journal} {\bibinfo  {journal}
  {Phys. Rev. A}\ }\textbf {\bibinfo {volume} {100}},\ \bibinfo {pages}
  {062118} (\bibinfo {year} {2019}{\natexlab{a}})}\BibitemShut {NoStop}%
\bibitem [{\citenamefont {Shukla}\ \emph {et~al.}(2023)\citenamefont {Shukla},
  \citenamefont {Modak},\ and\ \citenamefont {Mandal}}]{un_23}%
  \BibitemOpen
  \bibfield  {author} {\bibinfo {author} {\bibfnamefont {Namrata}\ \bibnamefont
  {Shukla}}, \bibinfo {author} {\bibfnamefont {Ranjan}\ \bibnamefont {Modak}},
  \ and\ \bibinfo {author} {\bibfnamefont {Bhabani~Prasad}\ \bibnamefont
  {Mandal}},\ }\bibfield  {title} {\enquote {\bibinfo {title} {Uncertainty
  relation for non-hermitian systems},}\ }\href {\doibase
  10.1103/PhysRevA.107.042201} {\bibfield  {journal} {\bibinfo  {journal}
  {Phys. Rev. A}\ }\textbf {\bibinfo {volume} {107}},\ \bibinfo {pages}
  {042201} (\bibinfo {year} {2023})}\BibitemShut {NoStop}%
\bibitem [{\citenamefont {Chitsazi}\ \emph {et~al.}(2017)\citenamefont
  {Chitsazi}, \citenamefont {Li}, \citenamefont {Ellis},\ and\ \citenamefont
  {Kottos}}]{exp_pt_1}%
  \BibitemOpen
  \bibfield  {author} {\bibinfo {author} {\bibfnamefont {Mahboobeh}\
  \bibnamefont {Chitsazi}}, \bibinfo {author} {\bibfnamefont {Huanan}\
  \bibnamefont {Li}}, \bibinfo {author} {\bibfnamefont {F.~M.}\ \bibnamefont
  {Ellis}}, \ and\ \bibinfo {author} {\bibfnamefont {Tsampikos}\ \bibnamefont
  {Kottos}},\ }\bibfield  {title} {\enquote {\bibinfo {title} {Experimental
  realization of floquet pt-symmetric systems},}\ }\href {\doibase
  10.1103/PhysRevLett.119.093901} {\bibfield  {journal} {\bibinfo  {journal}
  {Phys. Rev. Lett.}\ }\textbf {\bibinfo {volume} {119}},\ \bibinfo {pages}
  {093901} (\bibinfo {year} {2017})}\BibitemShut {NoStop}%
\bibitem [{\citenamefont {Biesenthal}\ \emph {et~al.}(2019)\citenamefont
  {Biesenthal}, \citenamefont {Kremer}, \citenamefont {Heinrich},\ and\
  \citenamefont {Szameit}}]{exp_pt_2}%
  \BibitemOpen
  \bibfield  {author} {\bibinfo {author} {\bibfnamefont {Tobias}\ \bibnamefont
  {Biesenthal}}, \bibinfo {author} {\bibfnamefont {Mark}\ \bibnamefont
  {Kremer}}, \bibinfo {author} {\bibfnamefont {Matthias}\ \bibnamefont
  {Heinrich}}, \ and\ \bibinfo {author} {\bibfnamefont {Alexander}\
  \bibnamefont {Szameit}},\ }\bibfield  {title} {\enquote {\bibinfo {title}
  {Experimental realization of pt-symmetric flat bands},}\ }\href {\doibase
  10.1103/PhysRevLett.123.183601} {\bibfield  {journal} {\bibinfo  {journal}
  {Phys. Rev. Lett.}\ }\textbf {\bibinfo {volume} {123}},\ \bibinfo {pages}
  {183601} (\bibinfo {year} {2019})}\BibitemShut {NoStop}%
\bibitem [{\citenamefont {Kremer}\ \emph {et~al.}(2019)\citenamefont {Kremer},
  \citenamefont {Biesenthal}, \citenamefont {Maczewsky}, \citenamefont
  {Heinrich}, \citenamefont {Thomale},\ and\ \citenamefont
  {Szameit}}]{exp_pt_3}%
  \BibitemOpen
  \bibfield  {author} {\bibinfo {author} {\bibfnamefont {Mark}\ \bibnamefont
  {Kremer}}, \bibinfo {author} {\bibfnamefont {Tobias}\ \bibnamefont
  {Biesenthal}}, \bibinfo {author} {\bibfnamefont {Lukas~J}\ \bibnamefont
  {Maczewsky}}, \bibinfo {author} {\bibfnamefont {Matthias}\ \bibnamefont
  {Heinrich}}, \bibinfo {author} {\bibfnamefont {Ronny}\ \bibnamefont
  {Thomale}}, \ and\ \bibinfo {author} {\bibfnamefont {Alexander}\ \bibnamefont
  {Szameit}},\ }\bibfield  {title} {\enquote {\bibinfo {title} {Demonstration
  of a two-dimensional pt-symmetric crystal},}\ }\href@noop {} {\bibfield
  {journal} {\bibinfo  {journal} {Nature communications}\ }\textbf {\bibinfo
  {volume} {10}},\ \bibinfo {pages} {435} (\bibinfo {year} {2019})}\BibitemShut
  {NoStop}%
\bibitem [{\citenamefont {Konotop}\ \emph
  {et~al.}(2016{\natexlab{b}})\citenamefont {Konotop}, \citenamefont {Yang},\
  and\ \citenamefont {Zezyulin}}]{exp_pt_4}%
  \BibitemOpen
  \bibfield  {author} {\bibinfo {author} {\bibfnamefont {Vladimir~V}\
  \bibnamefont {Konotop}}, \bibinfo {author} {\bibfnamefont {Jianke}\
  \bibnamefont {Yang}}, \ and\ \bibinfo {author} {\bibfnamefont {Dmitry~A}\
  \bibnamefont {Zezyulin}},\ }\bibfield  {title} {\enquote {\bibinfo {title}
  {Nonlinear waves in pt-symmetric systems},}\ }\href@noop {} {\bibfield
  {journal} {\bibinfo  {journal} {Reviews of Modern Physics}\ }\textbf
  {\bibinfo {volume} {88}},\ \bibinfo {pages} {035002} (\bibinfo {year}
  {2016}{\natexlab{b}})}\BibitemShut {NoStop}%
\bibitem [{\citenamefont {Ding}\ \emph {et~al.}(2021)\citenamefont {Ding},
  \citenamefont {Shi}, \citenamefont {Zhang}, \citenamefont {Shen},
  \citenamefont {Zhang},\ and\ \citenamefont {Zhang}}]{exp_pt_5}%
  \BibitemOpen
  \bibfield  {author} {\bibinfo {author} {\bibfnamefont {Liangyu}\ \bibnamefont
  {Ding}}, \bibinfo {author} {\bibfnamefont {Kaiye}\ \bibnamefont {Shi}},
  \bibinfo {author} {\bibfnamefont {Qiuxin}\ \bibnamefont {Zhang}}, \bibinfo
  {author} {\bibfnamefont {Danna}\ \bibnamefont {Shen}}, \bibinfo {author}
  {\bibfnamefont {Xiang}\ \bibnamefont {Zhang}}, \ and\ \bibinfo {author}
  {\bibfnamefont {Wei}\ \bibnamefont {Zhang}},\ }\bibfield  {title} {\enquote
  {\bibinfo {title} {Experimental determination of pt-symmetric exceptional
  points in a single trapped ion},}\ }\href {\doibase
  10.1103/PhysRevLett.126.083604} {\bibfield  {journal} {\bibinfo  {journal}
  {Phys. Rev. Lett.}\ }\textbf {\bibinfo {volume} {126}},\ \bibinfo {pages}
  {083604} (\bibinfo {year} {2021})}\BibitemShut {NoStop}%
\bibitem [{\citenamefont {Yu}\ \emph {et~al.}(2020)\citenamefont {Yu},
  \citenamefont {Meng}, \citenamefont {Tang}, \citenamefont {Xu}, \citenamefont
  {Wang}, \citenamefont {Yin}, \citenamefont {Ke}, \citenamefont {Liu},
  \citenamefont {Li}, \citenamefont {Yang}, \citenamefont {Chen}, \citenamefont
  {Han}, \citenamefont {Li},\ and\ \citenamefont {Guo}}]{exp_pt_6}%
  \BibitemOpen
  \bibfield  {author} {\bibinfo {author} {\bibfnamefont {Shang}\ \bibnamefont
  {Yu}}, \bibinfo {author} {\bibfnamefont {Yu}~\bibnamefont {Meng}}, \bibinfo
  {author} {\bibfnamefont {Jian-Shun}\ \bibnamefont {Tang}}, \bibinfo {author}
  {\bibfnamefont {Xiao-Ye}\ \bibnamefont {Xu}}, \bibinfo {author}
  {\bibfnamefont {Yi-Tao}\ \bibnamefont {Wang}}, \bibinfo {author}
  {\bibfnamefont {Peng}\ \bibnamefont {Yin}}, \bibinfo {author} {\bibfnamefont
  {Zhi-Jin}\ \bibnamefont {Ke}}, \bibinfo {author} {\bibfnamefont {Wei}\
  \bibnamefont {Liu}}, \bibinfo {author} {\bibfnamefont {Zhi-Peng}\
  \bibnamefont {Li}}, \bibinfo {author} {\bibfnamefont {Yuan-Ze}\ \bibnamefont
  {Yang}}, \bibinfo {author} {\bibfnamefont {Geng}\ \bibnamefont {Chen}},
  \bibinfo {author} {\bibfnamefont {Yong-Jian}\ \bibnamefont {Han}}, \bibinfo
  {author} {\bibfnamefont {Chuan-Feng}\ \bibnamefont {Li}}, \ and\ \bibinfo
  {author} {\bibfnamefont {Guang-Can}\ \bibnamefont {Guo}},\ }\bibfield
  {title} {\enquote {\bibinfo {title} {Experimental investigation of quantum
  pt-enhanced sensor},}\ }\href {\doibase 10.1103/PhysRevLett.125.240506}
  {\bibfield  {journal} {\bibinfo  {journal} {Phys. Rev. Lett.}\ }\textbf
  {\bibinfo {volume} {125}},\ \bibinfo {pages} {240506} (\bibinfo {year}
  {2020})}\BibitemShut {NoStop}%
\bibitem [{\citenamefont {Werner}(2008)}]{virial}%
  \BibitemOpen
  \bibfield  {author} {\bibinfo {author} {\bibfnamefont {F\'elix}\ \bibnamefont
  {Werner}},\ }\bibfield  {title} {\enquote {\bibinfo {title} {Virial theorems
  for trapped cold atoms},}\ }\href {\doibase 10.1103/PhysRevA.78.025601}
  {\bibfield  {journal} {\bibinfo  {journal} {Phys. Rev. A}\ }\textbf {\bibinfo
  {volume} {78}},\ \bibinfo {pages} {025601} (\bibinfo {year}
  {2008})}\BibitemShut {NoStop}%
\bibitem [{\citenamefont {Thomas}\ \emph {et~al.}(2005)\citenamefont {Thomas},
  \citenamefont {Kinast},\ and\ \citenamefont {Turlapov}}]{virial_exp1}%
  \BibitemOpen
  \bibfield  {author} {\bibinfo {author} {\bibfnamefont {J.~E.}\ \bibnamefont
  {Thomas}}, \bibinfo {author} {\bibfnamefont {J.}~\bibnamefont {Kinast}}, \
  and\ \bibinfo {author} {\bibfnamefont {A.}~\bibnamefont {Turlapov}},\
  }\bibfield  {title} {\enquote {\bibinfo {title} {Virial theorem and
  universality in a unitary fermi gas},}\ }\href {\doibase
  10.1103/PhysRevLett.95.120402} {\bibfield  {journal} {\bibinfo  {journal}
  {Phys. Rev. Lett.}\ }\textbf {\bibinfo {volume} {95}},\ \bibinfo {pages}
  {120402} (\bibinfo {year} {2005})}\BibitemShut {NoStop}%
\bibitem [{\citenamefont {Werner}\ and\ \citenamefont
  {Castin}(2006)}]{virial_exp2}%
  \BibitemOpen
  \bibfield  {author} {\bibinfo {author} {\bibfnamefont {F\'elix}\ \bibnamefont
  {Werner}}\ and\ \bibinfo {author} {\bibfnamefont {Yvan}\ \bibnamefont
  {Castin}},\ }\bibfield  {title} {\enquote {\bibinfo {title} {Unitary gas in
  an isotropic harmonic trap: Symmetry properties and applications},}\ }\href
  {\doibase 10.1103/PhysRevA.74.053604} {\bibfield  {journal} {\bibinfo
  {journal} {Phys. Rev. A}\ }\textbf {\bibinfo {volume} {74}},\ \bibinfo
  {pages} {053604} (\bibinfo {year} {2006})}\BibitemShut {NoStop}%
\bibitem [{\citenamefont {Shi}\ and\ \citenamefont
  {Sun}(2009)}]{shi2009recovering}%
  \BibitemOpen
  \bibfield  {author} {\bibinfo {author} {\bibfnamefont {T.}~\bibnamefont
  {Shi}}\ and\ \bibinfo {author} {\bibfnamefont {C.~P.}\ \bibnamefont {Sun}},\
  }\bibfield  {title} {\enquote {\bibinfo {title} {Recovering unitarity of lee
  model in biorthogonal basis},}\ }\href@noop {} {\  (\bibinfo {year}
  {2009})},\ \Eprint {http://arxiv.org/abs/0905.1771} {arXiv:0905.1771
  [hep-th]} \BibitemShut {NoStop}%
\bibitem [{\citenamefont {Kleefeld}(2009)}]{kleefeld2009construction}%
  \BibitemOpen
  \bibfield  {author} {\bibinfo {author} {\bibfnamefont {F.}~\bibnamefont
  {Kleefeld}},\ }\bibfield  {title} {\enquote {\bibinfo {title} {The
  construction of a general inner product in non-hermitian quantum theory and
  some explanation for the nonuniqueness of the c operator in pt quantum
  mechanics},}\ }\href@noop {} {\  (\bibinfo {year} {2009})},\ \Eprint
  {http://arxiv.org/abs/0906.1011} {arXiv:0906.1011 [hep-th]} \BibitemShut
  {NoStop}%
\bibitem [{\citenamefont {Ju}\ \emph {et~al.}(2019{\natexlab{b}})\citenamefont
  {Ju}, \citenamefont {Miranowicz}, \citenamefont {Chen},\ and\ \citenamefont
  {Nori}}]{PhysRevA.100.062118}%
  \BibitemOpen
  \bibfield  {author} {\bibinfo {author} {\bibfnamefont {Chia-Yi}\ \bibnamefont
  {Ju}}, \bibinfo {author} {\bibfnamefont {Adam}\ \bibnamefont {Miranowicz}},
  \bibinfo {author} {\bibfnamefont {Guang-Yin}\ \bibnamefont {Chen}}, \ and\
  \bibinfo {author} {\bibfnamefont {Franco}\ \bibnamefont {Nori}},\ }\bibfield
  {title} {\enquote {\bibinfo {title} {Non-hermitian hamiltonians and no-go
  theorems in quantum information},}\ }\href {\doibase
  10.1103/PhysRevA.100.062118} {\bibfield  {journal} {\bibinfo  {journal}
  {Phys. Rev. A}\ }\textbf {\bibinfo {volume} {100}},\ \bibinfo {pages}
  {062118} (\bibinfo {year} {2019}{\natexlab{b}})}\BibitemShut {NoStop}%
\bibitem [{\citenamefont {Tzeng}\ \emph {et~al.}(2021)\citenamefont {Tzeng},
  \citenamefont {Ju}, \citenamefont {Chen},\ and\ \citenamefont
  {Huang}}]{PhysRevResearch.3.013015}%
  \BibitemOpen
  \bibfield  {author} {\bibinfo {author} {\bibfnamefont {Yu-Chin}\ \bibnamefont
  {Tzeng}}, \bibinfo {author} {\bibfnamefont {Chia-Yi}\ \bibnamefont {Ju}},
  \bibinfo {author} {\bibfnamefont {Guang-Yin}\ \bibnamefont {Chen}}, \ and\
  \bibinfo {author} {\bibfnamefont {Wen-Min}\ \bibnamefont {Huang}},\
  }\bibfield  {title} {\enquote {\bibinfo {title} {Hunting for the
  non-hermitian exceptional points with fidelity susceptibility},}\ }\href
  {\doibase 10.1103/PhysRevResearch.3.013015} {\bibfield  {journal} {\bibinfo
  {journal} {Phys. Rev. Res.}\ }\textbf {\bibinfo {volume} {3}},\ \bibinfo
  {pages} {013015} (\bibinfo {year} {2021})}\BibitemShut {NoStop}%
\bibitem [{\citenamefont {Bartenstein}\ \emph {et~al.}(2004)\citenamefont
  {Bartenstein}, \citenamefont {Altmeyer}, \citenamefont {Riedl}, \citenamefont
  {Jochim}, \citenamefont {Chin}, \citenamefont {Denschlag},\ and\
  \citenamefont {Grimm}}]{density_exp_1}%
  \BibitemOpen
  \bibfield  {author} {\bibinfo {author} {\bibfnamefont {M.}~\bibnamefont
  {Bartenstein}}, \bibinfo {author} {\bibfnamefont {A.}~\bibnamefont
  {Altmeyer}}, \bibinfo {author} {\bibfnamefont {S.}~\bibnamefont {Riedl}},
  \bibinfo {author} {\bibfnamefont {S.}~\bibnamefont {Jochim}}, \bibinfo
  {author} {\bibfnamefont {C.}~\bibnamefont {Chin}}, \bibinfo {author}
  {\bibfnamefont {J.~Hecker}\ \bibnamefont {Denschlag}}, \ and\ \bibinfo
  {author} {\bibfnamefont {R.}~\bibnamefont {Grimm}},\ }\bibfield  {title}
  {\enquote {\bibinfo {title} {Crossover from a molecular bose-einstein
  condensate to a degenerate fermi gas},}\ }\href {\doibase
  10.1103/PhysRevLett.92.120401} {\bibfield  {journal} {\bibinfo  {journal}
  {Phys. Rev. Lett.}\ }\textbf {\bibinfo {volume} {92}},\ \bibinfo {pages}
  {120401} (\bibinfo {year} {2004})}\BibitemShut {NoStop}%
\bibitem [{den()}]{density_exp_2}%
  \BibitemOpen
  \href@noop {} {\ }\BibitemShut {NoStop}%
\bibitem [{\citenamefont {Bourdel}\ \emph {et~al.}(2004)\citenamefont
  {Bourdel}, \citenamefont {Khaykovich}, \citenamefont {Cubizolles},
  \citenamefont {Zhang}, \citenamefont {Chevy}, \citenamefont {Teichmann},
  \citenamefont {Tarruell}, \citenamefont {Kokkelmans},\ and\ \citenamefont
  {Salomon}}]{tof_1}%
  \BibitemOpen
  \bibfield  {author} {\bibinfo {author} {\bibfnamefont {T.}~\bibnamefont
  {Bourdel}}, \bibinfo {author} {\bibfnamefont {L.}~\bibnamefont {Khaykovich}},
  \bibinfo {author} {\bibfnamefont {J.}~\bibnamefont {Cubizolles}}, \bibinfo
  {author} {\bibfnamefont {J.}~\bibnamefont {Zhang}}, \bibinfo {author}
  {\bibfnamefont {F.}~\bibnamefont {Chevy}}, \bibinfo {author} {\bibfnamefont
  {M.}~\bibnamefont {Teichmann}}, \bibinfo {author} {\bibfnamefont
  {L.}~\bibnamefont {Tarruell}}, \bibinfo {author} {\bibfnamefont {S.~J. J.
  M.~F.}\ \bibnamefont {Kokkelmans}}, \ and\ \bibinfo {author} {\bibfnamefont
  {C.}~\bibnamefont {Salomon}},\ }\bibfield  {title} {\enquote {\bibinfo
  {title} {Experimental study of the bec-bcs crossover region in lithium 6},}\
  }\href {\doibase 10.1103/PhysRevLett.93.050401} {\bibfield  {journal}
  {\bibinfo  {journal} {Phys. Rev. Lett.}\ }\textbf {\bibinfo {volume} {93}},\
  \bibinfo {pages} {050401} (\bibinfo {year} {2004})}\BibitemShut {NoStop}%
\bibitem [{\citenamefont {Huang}\ \emph {et~al.}(2019)\citenamefont {Huang},
  \citenamefont {Lee}, \citenamefont {Zhang}, \citenamefont {Fei},\ and\
  \citenamefont {Wu}}]{weak_19}%
  \BibitemOpen
  \bibfield  {author} {\bibinfo {author} {\bibfnamefont {Minyi}\ \bibnamefont
  {Huang}}, \bibinfo {author} {\bibfnamefont {Ray-Kuang}\ \bibnamefont {Lee}},
  \bibinfo {author} {\bibfnamefont {Lijian}\ \bibnamefont {Zhang}}, \bibinfo
  {author} {\bibfnamefont {Shao-Ming}\ \bibnamefont {Fei}}, \ and\ \bibinfo
  {author} {\bibfnamefont {Junde}\ \bibnamefont {Wu}},\ }\bibfield  {title}
  {\enquote {\bibinfo {title} {Simulating broken pt-symmetric hamiltonian
  systems by weak measurement},}\ }\href {\doibase
  10.1103/PhysRevLett.123.080404} {\bibfield  {journal} {\bibinfo  {journal}
  {Phys. Rev. Lett.}\ }\textbf {\bibinfo {volume} {123}},\ \bibinfo {pages}
  {080404} (\bibinfo {year} {2019})}\BibitemShut {NoStop}%
\bibitem [{\citenamefont {Hatano}\ and\ \citenamefont
  {Nelson}(1996)}]{hatano1996localization}%
  \BibitemOpen
  \bibfield  {author} {\bibinfo {author} {\bibfnamefont {Naomichi}\
  \bibnamefont {Hatano}}\ and\ \bibinfo {author} {\bibfnamefont {David~R}\
  \bibnamefont {Nelson}},\ }\bibfield  {title} {\enquote {\bibinfo {title}
  {Localization transitions in non-hermitian quantum mechanics},}\ }\href@noop
  {} {\bibfield  {journal} {\bibinfo  {journal} {Physical review letters}\
  }\textbf {\bibinfo {volume} {77}},\ \bibinfo {pages} {570} (\bibinfo {year}
  {1996})}\BibitemShut {NoStop}%
\end{thebibliography}%
%%%%%%%%%
%\end{document}
%%%%%%%% Appendix A%%%%%%%%
\onecolumngrid
\appendix
\section{Four level System}

%A (4x4) matrix:

We consider a particular $4\times4$ non-Hermitian Hamiltonian as \cite{PhysRevA.103.062416},
$$H_{4\times{4}}=
\begin{pmatrix}{i\lambda}&{-1}&{0}&{0}\\ {-1}&{-i\lambda}&{-1}&{0}\\  {0} &{-1}&{i\lambda}&{-1}\\ {0} &{0}& {-1}&{-i\lambda}
\end{pmatrix}$$
%On finding the eigen values of this matrix, we observe the critical point as, $\lambda_{C_1}=\sqrt{\frac{3-\sqrt{5}}{2}}$, such that: for $\lambda<\lambda_{C_1}$ the spectrum is completely real and comes under unbroken phase. In the broken region region $(\lambda>\lambda_{c_1})$, it was found that there exists a point $\lambda_{C_2}=\sqrt{\frac{3+\sqrt{5}}{2}}$, such that in the region, $\lambda_{C_1}<\lambda<\lambda_{C_2}$, 1 pair of eigen values are real and 1 pair are imaginary, (which also comes under broken phase only) and for $\lambda>\lambda_{C_2}$ the spectrum is completely imaginary. This requires us to deduce G for all the three regions and verify the HFT.
The eigenvalues and right eigenfunctions for this system are calculated as \\ $E_1=-\sqrt{\frac{3-\sqrt{5}-2\lambda^2}{2}},\quad{E_2=\sqrt{\frac{3-\sqrt{5}-2\lambda^2}{2}}},\quad{E_3=-\sqrt{\frac{3+\sqrt{5}-2\lambda^2}{2}}},\quad{E_4=\sqrt{\frac{3+\sqrt{5}-2\lambda^2}{2}}}$ and 

%The right eigen vectors are,
$$|R_1\rangle=\begin{pmatrix}{u_{1_-}}\\ {v_1}\\ w_{1_+}\\ 1
\end{pmatrix},\quad|R_2\rangle=\begin{pmatrix}{u_{1_+}}\\ {v_1}\\w_{1_-}\\ 1
\end{pmatrix},\quad|R_3\rangle=\begin{pmatrix}{u_{2_-}}\\ {v_2}\\ w_{2_+}\\ 1
\end{pmatrix},\quad|R_4\rangle=\begin{pmatrix}{u_{2_+}}\\ {v_2}\\w_{2_-}\\ 1
\end{pmatrix}$$ 
%$$$$
%$$$$
%$$$$
where, \ba
u_{1_\pm}=\frac{1}{4} \left(2i\lambda\left(1+\sqrt{5}\right)\pm\left(\sqrt{2} +\sqrt{10}\right)\sqrt{3-\sqrt{5}-2\lambda^2}\right),\nonumber\\
{u_{2_\pm}}=\frac{1}{4} \left(2i\lambda\left(1-\sqrt{5}\right)\pm\left(\sqrt{2}-\sqrt{10}\right)\sqrt{3+\sqrt{5}-2\lambda^2}\right),\nonumber\\
{v_1}=\frac{\left(1-\sqrt{5}\right)}{2},\quad{v_2}=\frac{\left(1+\sqrt{5}\right)}{2},\nonumber\\
{w_{1_\pm}}=-i\lambda\pm\sqrt{\frac{3-\sqrt{5}-2\lambda^2}{2}},\quad{w_{2_\pm}}=-i\lambda\pm\sqrt{\frac{3+\sqrt{5}-2\lambda^2}{2}}.
\ea
%\begin{itemize}
% \item Unbroken $(\lambda<\lambda_{C_1})$
%\end{document}

We observe that if $\lambda<\lambda_{C_1}=\sqrt{\frac{3-\sqrt{5}}{2}},$ then all eigenvalues are real and if $\lambda_{C_1}<\lambda<\lambda_{C_2}=\sqrt{\frac{3+\sqrt{5}}{2}},$ then two of the eigenvalues are real and two are complex conjugate pair. For $\lambda>\lambda_{C_2},$ all four eigenvalues are complex. It can be shown that the system is in the unbroken phase for $\lambda<\lambda_{C_1}$.\\

We calculate the G metric in the unbroken phase using Eq.~\eqref{G_operator} as,

$$G^u=\frac{1}{10\left(1-3\lambda^2+\lambda^4\right)}
\begin{bmatrix}
 3-2\lambda^2&{i\lambda\left(3-2\lambda^2\right)}&-{\left(\lambda^2+1\right)}&{i\lambda\left(\lambda^2-4\right)}\\ 
{i\lambda\left(2\lambda^2-3\right)}&{2-3\lambda^2}&{i\left(\lambda^3+\lambda\right)}&-{\left(\lambda^2+1\right)}\\ 
-{\left(\lambda^2+1\right)}&-{i\left(\lambda^3+\lambda\right)}&{2-3\lambda^2}&{i\lambda\left(3-2\lambda^2\right)}\\ 
-{i\lambda\left(\lambda^2-4\right)}&-{\left(\lambda^2+1\right)}&{i\lambda\left(2\lambda^2-3\right)}&{3-2\lambda^2}
\end{bmatrix}$$
We explicitly check that $|R_1\rangle$ satisfies the modified HFT relation in Eq.~\eqref{hft_unbroken_2}:

\begin{equation}
\label{1revul}
    \frac{\partial{E_1}}{\partial\lambda}=\frac{\partial}{\partial{\lambda}}\left(-\sqrt{\frac{3-\sqrt{5}-2\lambda^2}{2}}\right)=\frac{\sqrt{2}\lambda}{\sqrt{3-\sqrt{5}-2\lambda^2}}
\end{equation}
We further calculate
\begin{equation}
\label{1revur}\Big\langle{R_1}\Big|G^u\frac{\partial{H_{4\times{4}}}}{\partial\lambda}\Big|R_1\Bigr\rangle=\frac{\sqrt{2}\lambda}{\sqrt{3-\sqrt{5}-2\lambda^2}},    
\end{equation}
to establish the modified HFT for the state $|R_1\rangle$, $$\boxed{\Big\langle{R_1}\Big|G^u\frac{\partial{H_{4\times{4}}}}{\partial\lambda}\Big|R_1\Bigr\rangle=\frac{\partial{E_1}}{\partial\lambda}}$$

In the similar way, it can be shown that other states also satisfy the modified HFT.\\

Now we consider the system in the broken phase, when $\lambda_{C_1}<\lambda<\lambda_{C_2}$. The eigenvalues in this region are, $E_1^{\prime}=-i\sqrt{\frac{-3+\sqrt{5}+2\lambda^2}{2}},\quad{E_2^{\prime}=i\sqrt{\frac{-3+\sqrt{5}+2\lambda^2}{2}}},\quad{E_3^{\prime}=-\sqrt{\frac{3+\sqrt{5}-2\lambda^2}{2}}},\quad{E_4^{\prime}=\sqrt{\frac{3+\sqrt{5}-2\lambda^2}{2}}}$\\

The G metric in this region of coupling $\lambda$ is calculated as

$$G^{b}_1=\frac{1}{10\left(1-3\lambda^2+\lambda^4\right)}
\begin{bmatrix}
   -\sqrt{5}&{-i\lambda\sqrt{5}}&{-\sqrt{5}\left(\lambda^2-1\right)}&-{i\lambda\sqrt{5}\left(\lambda^2-2\right)}\\
{i\lambda\sqrt{5}}&{2-\left(6+\sqrt{5}\right)\lambda^2+2\lambda^4}&{i\lambda\left(\lambda^2-1\right)}&g_{2,4}\\
{-\sqrt{5}\left(\lambda^2-1\right)}&-{i\lambda\left(\lambda^2-1\right)}&{-\sqrt{5}\lambda^2}&{-i\lambda\sqrt{5}}\\
{i\lambda\sqrt{5}\left(\lambda^2-2\right)}&g_{4,2}&{i\lambda\sqrt{5}}&g_{4,4}  
\end{bmatrix}$$
where, $g_{4,2}=g_{2,4}=-1+\left(3+2\sqrt{5}\right)\lambda^2-\left(1+\sqrt{5}\right)\lambda^4$\quad and\quad $g_{4,4}={3+\left(3+\sqrt{5}\right)\lambda^2\left(-3+\lambda^2\right)}$\\

We find that the states in this broken region also satisfy the modified HFT relation ( Eq.~\eqref{hft_broken_eqn}):

\begin{equation}
\label{1revubl}
    \frac{\partial{E_1^{\prime}}}{\partial\lambda}=\frac{\partial}{\partial{\lambda}}\left(-i\sqrt{\frac{-3+\sqrt{5}+2\lambda^2}{2}}\right)=-\frac{i\sqrt{2}\lambda}{\sqrt{-3+\sqrt{5}+2\lambda^2}}
\end{equation}
and,
\begin{equation}
\label{1revubr}\Big\langle{R_1}\Big|G^{b}_1\frac{\partial{H_{4\times{4}}}}{\partial\lambda}\Big|R_1\Bigr\rangle=-\frac{i\sqrt{2}\lambda}{\sqrt{-3+\sqrt{5}+2\lambda^2}}=\frac{\partial{E_1^{\prime}}}{\partial\lambda}    
\end{equation} implying that, $$\boxed{\Big\langle{R_1}\Big|G^{b}_1\frac{\partial{H_{4\times{4}}}}{\partial\lambda}\Big|R_1\Bigr\rangle=\frac{\partial{E_1^{'}}}{\partial\lambda}}$$

In the similar straightforward manner other states can be  shown to  satisfy the modified HFT in this broken region.\\

Finally, we consider the region $\lambda>\lambda_{C_2},$ where all eigenvalues are complex. The eigenvalues in this region are calculated as, $E_1^{\prime\prime}=-i\sqrt{\frac{-3+\sqrt{5}+2\lambda^2}{2}},\quad{E_2{\prime\prime}=i\sqrt{\frac{-3+\sqrt{5}+2\lambda^2}{2}}},\quad{E_3^{\prime\prime}=-i\sqrt{\frac{-3-\sqrt{5}+2\lambda^2}{2}}},\quad{E_4^{\prime\prime}=i\sqrt{\frac{-3-\sqrt{5}+2\lambda^2}{2}}}$.
 The G metric is constructed  as,
$$G^b_2=\frac{1}{10\left(1-3\lambda^2+\lambda^4\right)}
\begin{bmatrix}
 -3+2\lambda^2&{i\lambda\left(-3+2\lambda^2\right)}&{\left(\lambda^2+1\right)}&-{i\lambda\left(\lambda^2-4\right)}\\
{i\lambda\left(-2\lambda^2+3\right)}&{2-9\lambda^2+4\lambda^4}&-{i\left(\lambda^3+\lambda\right)}&-1+7\lambda^2-2\lambda^4\\
{\left(\lambda^2+1\right)}&{i\left(\lambda^3+\lambda\right)}&{-2+3\lambda^2}&{i\lambda\left(-3+2\lambda^2\right)}\\
{i\lambda\left(\lambda^2-4\right)}&-1+7\lambda^2-2\lambda^4&{i\lambda\left(-2\lambda^2+3\right)}&{3-16\lambda^2+6\lambda^4}
\end{bmatrix}$$
We now check below that  $|R_1\rangle$ still satisfies the modified HFT relation as given in Eq.~\eqref{hft_broken_eqn}:
The LHS of modified HFT is
\begin{equation}
\label{1revbl}
   \frac{\partial{E_1^{\prime\prime}}}{\partial\lambda}=\frac{\partial}{\partial{\lambda}}\left(-i\sqrt{\frac{-3+\sqrt{5}+2\lambda^2}{2}}\right)=-\frac{i\sqrt{2}\lambda}{\sqrt{-3+\sqrt{5}+2\lambda^2}}
\end{equation}
which is exactly equal to the RHS of the modified HFT. 
\begin{equation}
\label{1revbr}\Big\langle{R_1}\Big|G^b_2\frac{\partial{H_{4\times{4}}}}{\partial\lambda}\Big|R_1\Bigr\rangle=-\frac{i\sqrt{2}\lambda}{\sqrt{-3+\sqrt{5}+2\lambda^2}}=\frac{\partial{E_1^{\prime\prime}}}{\partial\lambda}  
\end{equation}
The same is true for all other states.
This establishes the verification of the modified HFT for $4\times4$ system in broken and unbroken phases.

\section{ HFT  for 2-d  anharmonic oscillator with non-Hermitian interaction}

We consider the 2-d anharmonic oscillator as described in Eq.~\eqref{pt_hamiltonian}. We choose the following  numerical values for different parameters, $\omega_x=1,\quad\omega_y=3,\quad\hbar=1,\quad{m}=1$ for the numerical computations.
The critical coupling $\lambda_c =\frac{m\omega_-^2}{2}$ becomes 4.

\begin{itemize}
    \item  For unbroken region $(\lambda<\lambda_c)$:
        
\begin{eqnarray}
\label{k0}&&k^*=k; \ \
C_1^*=C_1, \ C_2^*=C_2;\  \ \alpha_1^*=\alpha_1, \  \alpha_2^*=\alpha_2\\
&&\label{x^*}X^*=\sqrt{\frac{k+1}{2}}x+i\sqrt{\frac{k-1}{2}}y\\
&&\label{y^*}Y^* = -i\sqrt{\frac{k-1}{2}}x+\sqrt{\frac{k+1}{2}}y
\end{eqnarray}
\item For broken region $(\lambda>\lambda_c)$:

\begin{eqnarray}
    \label{kk*} && k^*=-k;\ \ C_1^*=C_2=A+iB; \ \ \alpha_1^*=\alpha_2, \ \ \alpha_2^*=\alpha_1\\ 
    \label{x**} &&X^*=i\sqrt{\frac{k-1}{2}}x-\sqrt{\frac{k+1}{2}}y\\
    \label{y**} && Y^*=\sqrt{\frac{k+1}{2}}x+i\sqrt{\frac{k-1}{2}}y
\end{eqnarray}
with \ba 
A=\frac{1}{2} \sqrt{\sqrt{\frac{\omega_-^4}{k_1^2}+\omega_+^4}+\omega_+^2};\ \ B=\frac{1}{2} \sqrt{\sqrt{\frac{\omega_-^4}{k_1^2}+\omega_+^4}-\omega_+^2};\ \ \mbox{and }
k_1=ik \ea
\end{itemize}

\begin{enumerate}
    \item Now for ground state $n_1=0,n_2=0,$ the energy eigenvalues $E_{0,0}$ are always  real for all values of $\lambda$, suggesting that there is no broken phase. We have calculated LHS and RHS of modified HFT analytically to show the equivalence. Using  Eqs.~\eqref{k0} to \eqref{y^*},  the  Eqs.~\eqref{right_eigen_vector_n1,n2} and \eqref{left_eigen_vector_n1,n2}  are rewritten as,
\begin{eqnarray}
\label{R0,0}
R_{0,0}&=& Ne^{-\frac{1}{4}\big[(C_1+C_2)(x^2+y^2)+(C_2-C_1)\big\{\frac{2i\lambda{kxy}}{\lambda_c}-k(x^2-y^2)\big\}\big]}\\
L_{0,0}&=& Ne^{-\frac{1}{4}\big[(C_1+C_2)(x^2+y^2)+(C_1-C_2)\left\{\frac{2i\lambda{kxy}}{\lambda_c}+k(x^2-y^2)\right\}\big]}\label{L0,0},
\end{eqnarray}

Now, we proceed to verify the modified HFT, using Eqs.~\eqref{R0,0} and \eqref{L0,0}
the LHS of Eq.~\eqref{hft_integral} implies,
$$\Big\langle\frac{\partial{H_{2d}}}{\partial\lambda}\Big\rangle_{G,(0,0)}=\frac{\int{{\Big(L_{0,0}^*}\Big)\frac{\partial{H_{2d}}}{\partial\lambda}\Big(R_{0,0}\Big)dxdy}}{\int{{\Big(L_{0,0}^*}\Big)\Big(R_{0,0}\Big)dxdy}}$$

\begin{eqnarray}
\label{lhs_of_27_0,0}
=\frac{|N|^2\int{ixye^{-\frac{m}{2\hbar}\big[(C_1+C_2)(x^2+y^2)+(C_2-C_1)\big\{\frac{2i\lambda{kxy}}{\lambda_c}-k(x^2-y^2)\big\}\big]}dxdy}}{|N|^2\int{e^{-\frac{m}{2\hbar}\big[(C_1+C_2)(x^2+y^2)+(C_2-C_1)\big\{\frac{2i\lambda{kxy}}{\lambda_c}-k(x^2-y^2)\big\}\big]}dxdy}}\nonumber\\
=-\frac{\lambda  \left(\lambda ^2-4 \sqrt{\lambda ^2+9}+4\right) \sqrt{5-\sqrt{16-\lambda ^2}}}{\left(16-\lambda ^2\right) \left(\lambda ^2+8\right) \sqrt{\lambda ^2+9}}-\frac{\lambda  \left(\lambda ^2-4 \sqrt{\lambda ^2+9}+4\right) \sqrt{5-\sqrt{16-\lambda ^2}}}{4 \sqrt{16-\lambda ^2} \left(\lambda ^2+8\right) \sqrt{\lambda ^2+9}}\nonumber\\
+\frac{\lambda  \left(\lambda ^2-4 \sqrt{\lambda ^2+9}+4\right) \sqrt{\sqrt{16-\lambda ^2}+5}}{4 \sqrt{16-\lambda ^2} \left(\lambda ^2+8\right) \sqrt{\lambda ^2+9}}-\frac{\lambda  \left(\lambda ^2-4 \sqrt{\lambda ^2+9}+4\right) \sqrt{\sqrt{16-\lambda ^2}+5}}{\left(16-\lambda ^2\right) \left(\lambda ^2+8\right) \sqrt{\lambda ^2+9}}
\end{eqnarray}

RHS of Eq.~\eqref{hft_integral} for the $(0,0)$ state is,

\begin{eqnarray}
\label{rhs_of_27_0,0}
\frac{\partial{E_{0,0}}}{\partial\lambda}=\frac{\partial}{\partial\lambda}\Big(\frac{1}{2}{C_1}+\frac{1}{2}{C_2}\Big)=\frac{\lambda \sqrt{\sqrt{\lambda^2+9}-5}}{2\sqrt{2}\sqrt{\lambda^2-16} \sqrt{\lambda^2+9}}   
\end{eqnarray}
The RHS of the Eqs.~\eqref{lhs_of_27_0,0} and \eqref{rhs_of_27_0,0} are same for all values of $\lambda$, as shown in Fig.~\ref{fig:GS}, indicating that modified HFT is valid for the ground state $n_1=n_2=0$, for all values of $\lambda$.

\begin{figure}[h]
    \centering \includegraphics[width=0.6\textwidth]{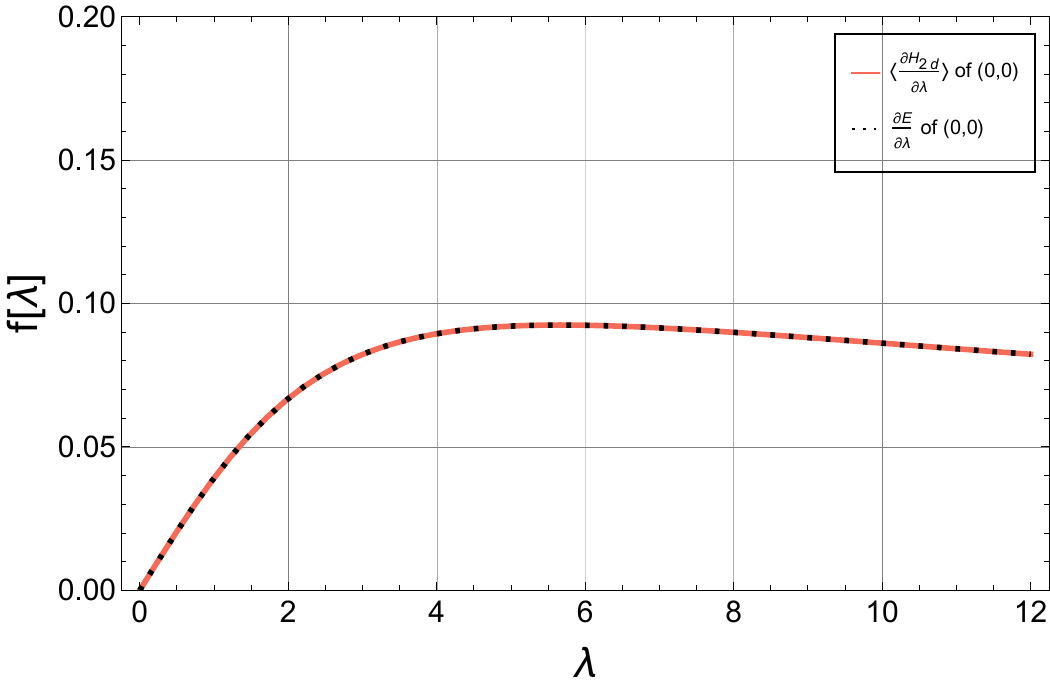}
    \caption{Comparison between $\Big\langle\frac{\partial{H_{2d}}}{\partial\lambda}\Big\rangle_{G,(0,0)}$ and $\frac{\partial{E_{0,0}}}{\partial\lambda}$, represented by  the  solid and the dotted lines  respectively.}
    \label{fig:GS}
\end{figure}

\item First excited state: $n_1=1,n_2=0$.

\begin{itemize}
\item {\bf Unbroken phase:} $\lambda<\lambda_c$

Using the relations in Eqs.~\eqref{k0} to \eqref{y^*}, 
the right and the left eigenvectors are derived as,
\begin{eqnarray}
\label{R1,0}
R_{1,0}&=&Ne^{-\frac{1}{4}\big[(C_1+C_2)(x^2+y^2)+(C_2-C_1)\big\{\frac{2i\lambda{kxy}}{\lambda_c}-k(x^2-y^2)\big\}\big]} 2\alpha_1\left(\sqrt{\frac{k+1}{2}}x-i\sqrt{\frac{k-1}{2}y}\right)\\
L_{1,0}&=&Ne^{-\frac{1}{4}\big[(C_1+C_2)(x^2+y^2)+(C_1-C_2)\big\{\frac{2i\lambda{kxy}}{\lambda_c}+k(x^2-y^2)\big\}\big]} 2\alpha_1\left(\sqrt{\frac{k+1}{2}}x+i\sqrt{\frac{k-1}{2}y}\right)\label{L1,0} 
\end{eqnarray}

with the common eigenvalue, $\quad\quad{E_{1,0}^*}=E_{1,0}=(\frac{3}{2}){C_1}+(\frac{1}{2}){C_2}$.
Hence, the LHS of modified HFT  is,
$$\Big\langle\frac{\partial{H_{2d}}}{\partial\lambda}\Big\rangle_{G,(1,0)}=\frac{\int{{\Big(L_{1,0}^*}\Big)\frac{\partial{H_{2d}}}{\partial\lambda}\Big(R_{1,0}\Big)dxdy}}{\int{{\Big(L_{1,0}^*}\Big)\Big(R_{1,0}\Big)dxdy}}$$
\begin{eqnarray}
\label{lhs_of_27_1,0}
=\frac{|N|^2\int{ixyc_1\left(\sqrt{\frac{k+1}{2}}x-i\sqrt{\frac{k-1}{2}}y\right)^2e^{-\frac{1}{2}\big[(C_1+C_2)(x^2+y^2)+(C_2-C_1)\big\{\frac{2i\lambda{kxy}}{\lambda_c}-k(x^2-y^2)\big\}\big]}dxdy}}{|N|^2\int{c_1\left(\sqrt{\frac{k+1}{2}}x-i\sqrt{\frac{k-1}{2}}y\right)^2e^{-\frac{1}{2}\big[(C_1+C_2)(x^2+y^2)+(C_2-C_1)\big\{\frac{2i\lambda{kxy}}{\lambda_c}-k(x^2-y^2)\big\}\big]}dxdy}}
\end{eqnarray}

And the RHS of modified HFT is,
\begin{eqnarray}
\label{rhs_of_27_1,0}
\frac{\partial{E_{1,0}}}{\partial\lambda}=\frac{\partial}{\partial\lambda}\Big(\frac{3}{2}{C_1}+\frac{1}{2}{C_2}\Big)=\frac{3 \lambda }{8 \sqrt{2\left(1-\frac{\lambda ^2}{16}\right)\left(10-8 \sqrt{1-\frac{\lambda ^2}{16}}\right)}}-\frac{\lambda }{8 \sqrt{2\left(1-\frac{\lambda ^2}{16}\right)\left(10+8 \sqrt{1-\frac{\lambda ^2}{16}}\right)}}  
\end{eqnarray}

The expressions in the RHS of the Eqs.~\eqref{lhs_of_27_1,0} and \eqref{rhs_of_27_1,0} in the domain $\lambda<\lambda_c(=4)$ are equal as shown in Fig.~\ref{fig:1,0*abo}, indicating that the modified HFT is valid for the state $(n_1=1,n_2=0)$ in the unbroken case.

\item {\bf Broken phase:} $\lambda>\lambda_c$

Using the relations in Eqs.~\eqref{kk*} to \eqref{y**}, the right and the left eigenvectors are calculated  along with their eigenvalues as,
\begin{eqnarray}
\label{broken_REV_1,0}
R_{1,0}=Ne^{-\frac{1}{4}\big[2A(x^2+y^2)+2iB\big\{\frac{2i\lambda{kxy}}{\lambda_c}-k(x^2-y^2)\big\}\big]} 2\alpha_1\left(\sqrt{\frac{k+1}{2}}x-i\sqrt{\frac{k-1}{2}}y\right)
\end{eqnarray} 
\begin{eqnarray}
\label{1,0_broken_ev}
{E_{1,0}}=\frac{3}{2}{C_1}+\frac{1}{2}{C_2}=\frac{3}{2}{(A-iB)}+\frac{1}{2}{(A+iB)}=2A-iB    
\end{eqnarray}
And,
\begin{eqnarray}
\label{broken_LEV_1,0}
L_{1,0}=Ne^{-\frac{1}{4}\big[2A(x^2+y^2)-2iB\big\{\frac{2i\lambda{kxy}}{\lambda_c}+k(x^2-y^2)\big\}\big]} 2\alpha_1^*\left(i\sqrt{\frac{k-1}{2}}x-\sqrt{\frac{k+1}{2}}y\right)
\end{eqnarray}
with eigenvalue $E_{1,0}^*$.
 The LHS of the modified HFT is calculated as,

\begin{eqnarray}
\Big\langle\frac{\partial{H_{2d}}}{\partial\lambda}\Big\rangle_{G,(1,0)}&=&\frac{\int{{\Big(L_{1,0}^*}\Big)\frac{\partial{H_{2d}}}{\partial\lambda}\Big(R_{1,0}\Big)dxdy}}{\int{{\Big(L_{1,0}^*}\Big)\Big(R_{1,0}\Big)dxdy}}\nonumber \\
\label{lhs_of_27_1,0*}
&=&\frac{|N|^2\int{ixyc_1\left(\sqrt{\frac{k+1}{2}}x-i\sqrt{\frac{k-1}{2}}y\right)^2e^{-\frac{1}{2}\big[2A(x^2+y^2)+i2B\big\{\frac{2i\lambda{kxy}}{\lambda_c}-k(x^2-y^2)\big\}\big]}dxdy}}{|N|^2\int{c_1\left(\sqrt{\frac{k+1}{2}}x-i\sqrt{\frac{k-1}{2}}y\right)^2e^{-\frac{1}{2}\big[2A(x^2+y^2)+i2B\big\{\frac{2i\lambda{kxy}}{\lambda_c}-k(x^2-y^2)\big\}\big]}dxdy}}
\end{eqnarray}

RHS of modified HFT is,

\begin{eqnarray}
\label{rhs_of_27_1,0*}
&&\frac{\partial{E_{1,0}}}{\partial\lambda}
=\frac{\partial}{\partial\lambda}\Big(2A-iB\Big)\nonumber\\
&=&\frac{2 \lambda }{\sqrt{64 \left(\frac{\lambda ^2}{16}-1\right)+100} \sqrt{\sqrt{64 \left(\frac{\lambda ^2}{16}-1\right)+100}+10}}-\frac{i\lambda }{\sqrt{64 \left(\frac{\lambda ^2}{16}-1\right)+100} \sqrt{\sqrt{64 \left(\frac{\lambda ^2}{16}-1\right)+100}-10}}
\end{eqnarray}

The imaginary and the real parts of the relations given by Eqs.~\eqref{lhs_of_27_1,0*} and \eqref{rhs_of_27_1,0*} over the broken region are shown in Fig.~\ref{fig:1,0*re_im}. We have also shown the real and imaginary parts of Eq.~\eqref{hft_integral} for $(2,0)$ state in the same figure in order to compare it with the $(1,0)$ state. We find that the curves representing the LHS and RHS of Eq.~\eqref{hft_integral} completely overlap each other for both $(1,0)$ and $(2,0)$ states, indicating the validity of the modified HFT for these states as well.
%%%%%%%

\begin{figure}[h!]  \centering\includegraphics[width=0.7\textwidth]{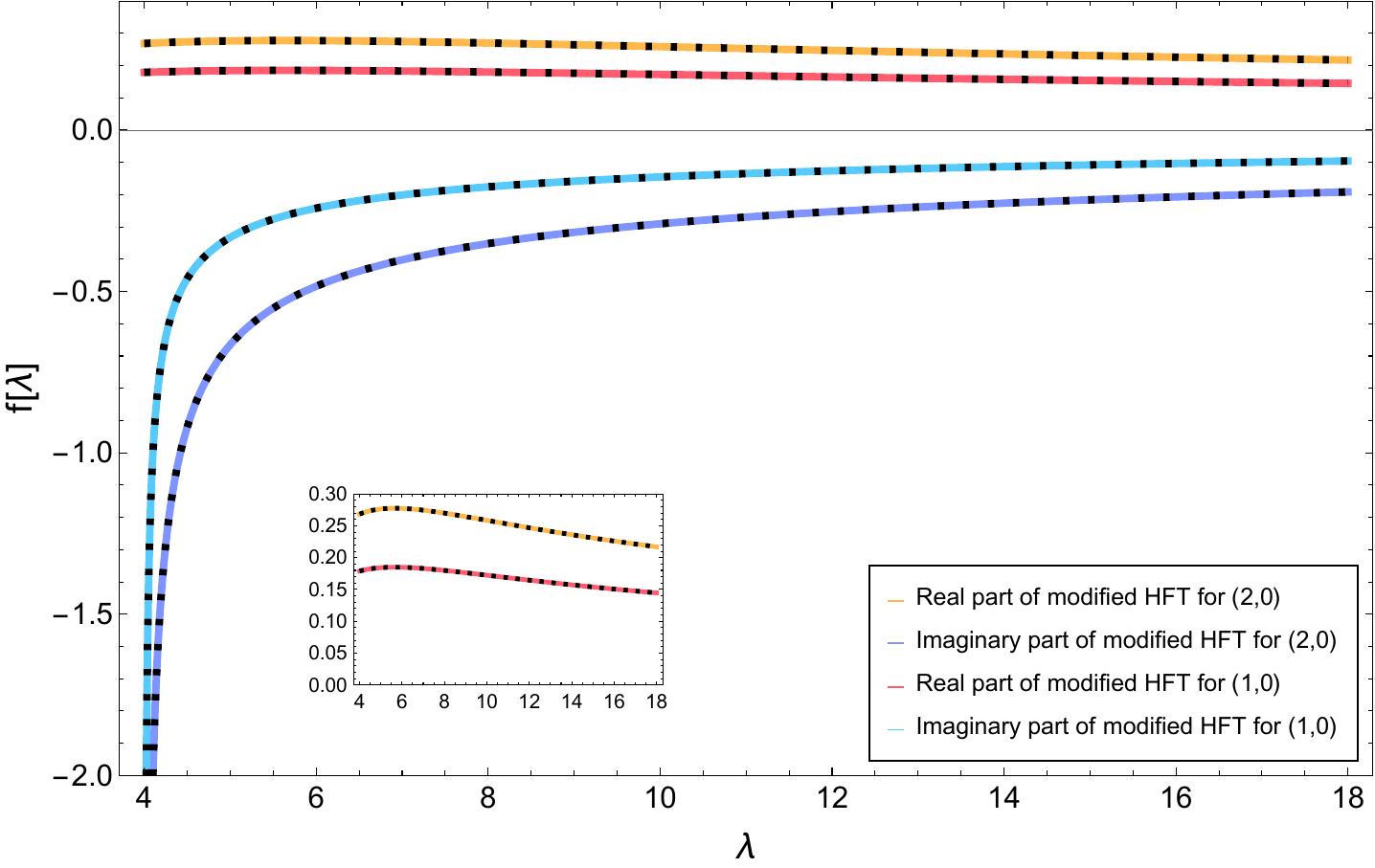}
    \caption{Comparison between real and imaginary parts of $\Big\langle\frac{\partial{H_{2d}}}{\partial\lambda}\Big\rangle_{G,(1,0)}$ and $\Big\langle\frac{\partial{H_{2d}}}{\partial\lambda}\Big\rangle_{G,(2,0)}$ with that of $\frac{\partial{E_{1,0}}}{\partial\lambda}$ and $\frac{\partial{E_{2,0}}}{\partial\lambda}$ in the  broken region. Colored solid lines and black dotted lines represent the LHS and RHS of modified HFT respectively. }
    \label{fig:1,0*re_im}
\end{figure}

%Thus, HFT is valid for non-hermitian PT symmetric Hamiltonian.
\end{itemize}
\end{enumerate}

%%%%%%%%%%%%%%%

\section{HFT for non-PT invariant non-Hermitian system}

We consider here a simple 1-d  Hamiltonian( Eq.~\eqref{NH-NPT-symmetric})representing non-PT symmetric non-hermitian system 

\be
\label{NH-NPTNO}
H_{1d}=\frac{p^2}{2}+\frac{1}{2}\Omega^2x^2 .
\ee

where, $\Omega=\omega_1+i\omega_2$ and $\omega_{1,2}$ are real and $\omega_{1,2}\ne0.$

We intend to verify the modified HFT
\be
\label{hft-NH1DNO}
\Big\langle\frac{\partial{H_{1d}}}{\partial\lambda}\Big\rangle_{G,(n)}=\frac{\partial{E_n}}{\partial\lambda}
\ee 
for the above system for an arbitrary $n^{th}$ state with 3 different choices of $\lambda:$

\begin{enumerate}
\item First choice, $\lambda=\omega_1$:
Considering the relations in equations from Eqs.~\eqref{x_rel} to \eqref{a_adag}, we can write,
\be
\label{Non-LHS}
\Big\langle\frac{\partial{H_{1d}}}{\partial\omega_1}\Big\rangle_{G,(n)}=\langle{\bar{n}}|\Omega{x^2}|n\rangle=\frac{1}{2}\langle{\bar{n}}|{a^+}^2+a^2+a^+a+aa^+|n\rangle=\frac{1}{2}\langle{\bar{n}}|2n+1|n\rangle=\left(n+\frac{1}{2}\right)
\ee

And it is straight-forward to check that,
\be
\label{Non-H}
\boxed{\frac{\partial{E_n}}{\partial\omega_1}
=\left(n+\frac{1}{2}\right)=\Big\langle\frac{\partial{H_{1d}}}{\partial\omega_1}\Big\rangle_{G,(n)}}
\ee

\item Second choice, $\lambda=\omega_2$:

Here too it is straight forward to check,
\be
\label{Non-LHS1}
\Big\langle\frac{\partial{H_{1d}}}{\partial\omega_2}\Big\rangle_{G,(n)}=\langle{\bar{n}}|i\Omega{x^2}|n\rangle=\frac{i}{2}\langle{\bar{n}}|{a^+}^2+a^2+a^+a+aa^+|n\rangle=\frac{i}{2}\langle{\bar{n}}|2n+1|n\rangle=i\left(n+\frac{1}{2}\right)
\ee

Again,
\be
\label{Non-H1}
\boxed{\frac{\partial{E_n}}{\partial\omega_2}
=i\left(n+\frac{1}{2}\right)=\Big\langle\frac{\partial{H_{1d}}}{\partial\omega_2}\Big\rangle_{G,(n)}}
\ee

\item Third choice, $\lambda=\Omega$:

Similarly for this choice,
\be
\label{Non-LHS2}
\Big\langle\frac{\partial{H_{1d}}}{\partial\Omega}\Big\rangle_{G,(n)}=\langle{\bar{n}}|\Omega{x^2}|n\rangle=\frac{1}{2}\langle{\bar{n}}|{a^+}^2+a^2+a^+a+aa^+|n\rangle=\frac{1}{2}\langle{\bar{n}}|2n+1|n\rangle=\left(n+\frac{1}{2}\right)
\ee

Again,
\be
\label{Non-H2}
\boxed{\frac{\partial{E_n}}{\partial\Omega}=\left(n+\frac{1}{2}\right)=\Big\langle\frac{\partial{H_{1d}}}{\partial\Omega}\Big\rangle_{G,(n)}}
\ee
\end{enumerate}
From Eqs.~\eqref{Non-H}, \eqref{Non-H1} and \eqref{Non-H2}, we see that the modified HFT is valid in case of non-PT invariant non-Hermitian system as well.

\end{document}